# Quasi-molecular electron promotion beyond the 1sσ and 2pπ channels in slow collisions of He$^{2+}$ and He

L. Ph. H. Schmidt, M. Schöffler, C. Goihl, T. Jahnke, H. Schmidt-Böcking,
R. Dörner

*Institut für Kernphysik, Goethe-Universität, 60438 Frankfurt am Main, Germany*

The electron emission pattern of transfer-ionization in collisions of He$^{2+}$ with He was investigated for impact velocities between 0.53 a.u. and 0.77 a.u. (7 keV/u – 15 keV/u) employing recoil-ion momentum spectroscopy. This process is known to be dominated by the promotion of the 2pπ molecular orbital into the continuum which results in "banana" shaped areas of high electron momentum densities extending from the target to the projectile in velocity space in the collision plain. Asymmetries are explained by a coherent superposition of the 2pπ channel of the quasi molecular promotion with the 1sσ channel. Here we report on an additional contribution of higher angular momentum molecular states for close collisions which emerge at smaller impact velocities. They show up as highly structured electron emission patterns in the plane perpendicular to the direction of impact.

## I. Introduction

The collision of an ion with an atom can lead to the ejection of one or more electrons from the target to the continuum. For fast collisions, where the velocity of the projectile is faster than the typical velocity of the bound electrons, distinct features in the electron momentum distributions (such as the electron capture to the continuum (ECC) and the binary encounter peak) occur and are well understood nowadays. In slow collisions a binary encounter between the projectile and an electron in the target atom typically cannot transfer sufficient energy to knock out the electron from its orbital. A first suggestion for a mechanism which can still lead to electron ejection in such slow collisions dates back to the 1980s: Olson [1] found in classical trajectory calculations continuum electrons stranded between the projectile and target ion where the attraction of the two nuclei is at balance. These so called saddle point electrons are emitted with approximately half of the projectile velocity in the direction of the ion impact.

This classical picture of an ionization process via promotion on the saddle point has then been confirmed and refined in quantum mechanical approaches. E.g. close coupling calculations successfully describe the electron transfer for slow collisions [2],. To represent the electron wave function in such close coupling calculations, a basis set containing atomic wave functions centered at both nuclei can be used. At impact velocities much below 1 a.u. it is, however, appropriate to combine the atomic states to a quasi-molecular basis [3] and then use the close coupling approach to calculate transitions between theses states. For example p-H collisions can be described by the molecular electronic states of $H_2^+$. Within this basis the initial and final state where the electron is located at one or the other nucleus is a superposition of the two lowest molecular states $1s\sigma_g$ and $2p\sigma_u$. The notation of the molecular states refers to the limit of united atoms. The actual location of the electron depends on the phase between these states. Therefore an electron transfer is simply induced by a phase shift between the gerade and ungerade molecular states which can emerge because the binding energies of these states differ at small internuclear distances.

When the projectile is passing the target the internuclear axis between both rotates rapidly. This rotation effectively couples the $2p\sigma_u$ and $2p\pi_u$ states. In the asymptote of large internuclear distances this leads to electrons excited to the 2p atomic state. In addition to this coupling by rotation the change of the internuclear distance causes a coupling between states of the same symmetry as $1s\sigma_g$ and $3s\sigma_g$ [4]. These radial couplings result from an inadequate electron momentum space representation of the dynamical problem by adiabatic states in direction of the internuclear vector. Therefore the radial coupling gets larger if the electron is excited to higher molecular orbitals states and thereby the momentum spread of the states gets small with respect to the error of the adiabatic description.

In contrast to the quasi-molecular basis states the atomic wave functions can be easily adapted to the impact velocity but these wave functions fail to give a good description of the effects appearing at the closest approach of the nuclei. To combine the advantages of a molecular basis and a two center atomic basis the triple-center treatment was developed [5]. Winter and Lin [6] applied this method to the ionization-channel of p-H collisions aiming to represent the electron momentum space of electrons emitted with low energy with respect to the saddle of the nuclear coulomb potential, as well as the bound states with sufficient precision. Triple-center calculations are also available for $He^{2+}$ on He collisions [7] but electron emission patterns have not been reported for this collision system.

A pertinent problem for describing electron emission to the continuum in a slow collision within any close coupling scheme is, that the electrons reach the continuum by "climbing up a ladder" of infinitely many Rydberg states via radial couplings between these states. To avoid the explicit treatment of these radial couplings through the whole Rydberg series of molecular states the Hidden Crossing Method (HC) [8, 9, 10] uses molecular pseudo states which merge an infinite number of adiabatic molecular states of the same symmetry. This method describes the contributions of several molecular symmetries to the ionization cross section, but it is still not capable to produce correct momentum distributions of the continuum electrons.

This problem has been solved by Schultz, Ovchinnikov and Macek in 2014 by switching from the HC method to a regularized lattice (RL) representation of the electronic wave function at a certain internuclear separation after the closest approach. [11]. The method which describes the transfer ionization (TI) $^{proj.}He^{2+}$ + He ➔ $^{proj.}He^{+}$ + $He^{2+}$ + $e^{-}$ was abbreviated by 2eHC-RLTDSE [12]. The RL calculations are done for only one active electron and a single impact parameter. But because the relevant rotational couplings appear during the first stage of the calculation using the 2eHC approach the resulting electron emission pattern still include the features related to two active electrons and a coherent superposition of different nuclear trajectories representing a specific momentum transfer.

Schmidt et al. have shown [13] that the main features of the electron emission pattern of this reaction at impact energies higher than 10 keV/u can be explained by a superposition of the quasi-molecular promotion of the $1s\sigma_g$ and $2p\pi_u$ states. Several other collision systems show similar behavior [14, 15 16] because a rotational coupling of a single active electron can not populate states of higher angular momentum. For an extension of these quasi-molecular features in the electron momentum distributions to higher impact energies see [17].

The key difference between single ionization and TI is that the participation of both electrons in TI makes it more likely to transfer an angular momentum of $2\hbar$ due to rotational coupling from the nuclear motion to the electronic state instead of only $1\hbar$ in a single electron process. Recently Schmidt et al. reported [12] that the electron emission pattern of the reaction

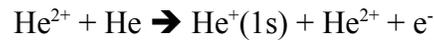

$$He^{2+} + He ➔ He^{+}(1s) + He^{2+} + e^{-}$$

at 10 keV/u projectile energy shows the signatures of $3d\delta_u$ and $2s\sigma_g$ states. The superposition of these two channels leads to creation of free vortices in the wave function of the emitted electron. The vortices only show up if reactions with a specific nuclear momentum exchange are selected.

In this paper we give a more detailed and more complete discussion of the extended series of experiments reported in [12]. We present measurements at six impact energies between 7 keV/u and 15 keV/u. The datasets are subdivided into many regions of internuclear momentum exchange. This comprehensive study of impact velocities and scattering angles led to the discovery of further and so far hidden channels of quasi-molecular electron promotion to the continuum.

## II Experiment

The experiment was performed using a penning ion source at the Institut für Kernphysik of the Goethe-University Frankfurt. A $^4$He$^{2+}$ beam was accelerated with voltages between 14 keV and 30 keV. A magnetic mass-to-charge ratio selection was sufficient to separate the He$^{2+}$ ions from H$_2^+$. Before entering the reaction chamber the beam was collimated to a diameter less than 0.5 mm and a divergence less than 0.5 mrad.

The He$^{2+}$ ion beam was crossed at 90 deg. with a super sonic Helium gas jet. The super sonic Helium beam was produced by expanding gas at 16 bar through a 30 μm nozzle into a vacuum of 0.02 mbar. The nozzle was cooled to 140 K. Due to the expansion the internal temperature of the gas dropped to about 100 mK. With two differentially pumped skimmers (with 0.3 mm aperture opening each) we cut out a narrow and internally cold Helium beam which had a diameter of 1 mm at the interaction region with the ion beam. We used a double-stage differentially pumped gas jet beam dump to keep the back-pressure inside the reaction chamber below $10^{-8}$ mbar while the local density of Helium atoms within the gas beam corresponds to a pressure above $10^{-5}$ mbar.

After leaving the reaction region the projectile beam was charge-state selected by an electrostatic deflection scheme as sketched in Fig. 1(a). The projectiles which captured one or two electrons were detected by a multi channel plate detector (MCP) with delay line position readout [18] placed about 1.3 m behind the reaction region while the He$^{2+}$ projectiles (that did not react with the target beam) were dumped in a Faraday cup. The projectile detector mainly provides the information on the charge state and gives a time reference for the time-of-flight measurements of the recoil-ion and the electron. However, the time information has an uncertainty of about 1 ns caused by the projectile time of flight through the reaction region. This is the major limitation of the momentum spectroscopy of the emitted electron and the ionized target atoms which are detected in coincidence with the projectiles by using a reaction microscope [19].

The Helium recoil-ions and the electrons were extracted to opposite directions by a weak electric field of about 0.15 V/mm perpendicularly to the plane spanned by the two crossed beams [see Fig. 1(a)]. They were detected by two further position and time sensitive MCP detectors.

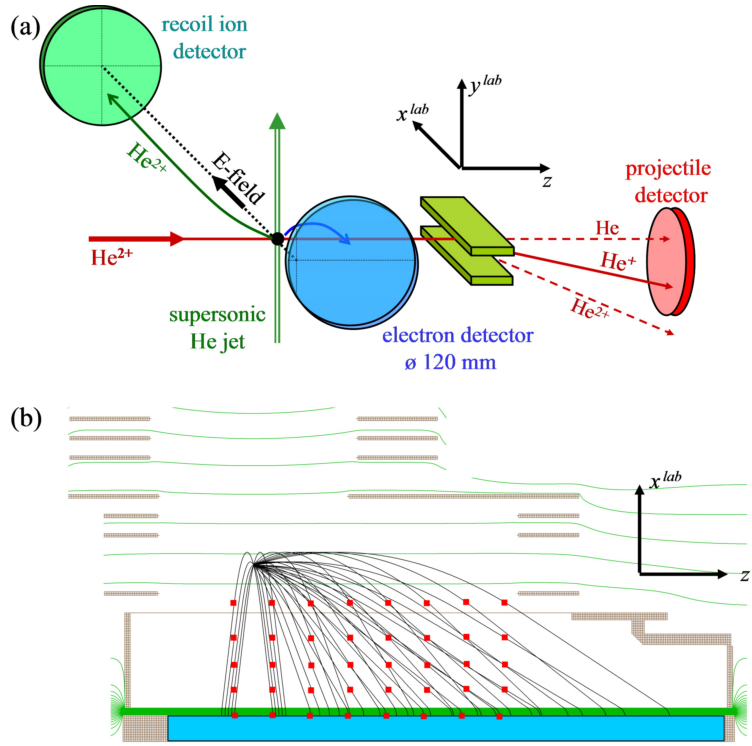

Fig 1: Experimental setup: (a) Arrangement of the three detectors. (b) Simulated trajectories of electrons with momenta $p_x=$ -0.2, -0.1, 0, 0.1 , 0.2 a.u. and $p_z=$ -0.05, 0.05, 0.15, … , 0.65 a.u.. The red squares show the electron positions 50 ns after the reaction time. The electrons with a momentum of 0.2 a.u. in direction of the detector already hit the detector.

The electron arm of the momentum analyzer consists of an acceleration region of 13 mm length followed by a drift region of 26 mm separated by a mesh. After the drift the electrons are accelerated by a high field to the micro channel plate detector with a diameter of 124 mm. The centre of this detector is shifted by 55 mm from the spectrometer axis in direction of the projectile beam in order to spread the spatial distribution of electrons that occur at velocities between zero and the projectile velocity over nearly the whole active surface of this detector [see Fig 1(b)]. In the detector plane we achieved an electron momentum resolution of 0.01 a.u. mainly limited by the size of the reaction region of 1 mm (a.u. denotes atomic units: $m_e =$ e $= \hbar = c/137 = 1$). In direction of the electric field the electron momentum was calculated from the time-of-flight resulting in a momentum resolution of about 0.02 a.u. (FWHM), which mainly arises from the uncertainty of the timing reference obtained by the projectile detection.

The recoil-ions passed through a system of electrostatic lenses and a drift region which minimizes the momentum uncertainty caused by the size of the reaction region. The electric field in the reaction region was optimized for the electron arm of the spectrometer which requires a homogeneous and very low field. The low electric field impacts the design of the recoil-ion arm: low electric fields cause very long ion times-of-flight. Thus the recoil-ion distribution expands spatially to more than 40 mm before a first electrostatic lens can be placed. Therefore a long recoil-ion drift region became necessary to achieve time-of-flight focusing yielding a total length of the recoil arm of about 1 m. Therein the ions' spatial distribution expands to 150 mm diameter perpendicular to the spectrometer axis before it can be refocused onto the detector of 80 mm diameter. The spectrometer voltage settings have been slightly changed between the measurements at different impact velocities. For all settings the recoil-ion momentum resolution was better than 0.2 a.u.

Both electron and recoil-ion momentum distributions are rotational symmetric with respect to the direction of the incoming beam which defines the z axis of our coordinate system. These symmetries can be used to adjust the calibration factors and offsets in the transversal plane. In the laboratory frame of reference the two axes perpendicular to the direction of impact are defined by the electric field direction ($x^{lab}$ axis) and the propagation direction of the target gas beam ($y^{lab}$ axis).

In the transverse direction the momentum transfer to the projectile and the projectile scattering angle are inferred from the measured momenta of the recoil-ion and electron using momentum conservation. Additionally we measured the scattering angle directly on the projectile detector but the corresponding projectile transversal momentum resolution was only about 5 a.u. (FWHM). Nevertheless, most of the background from statistical coincidences between projectiles and recoil-ions can be eliminated by considering only those events which fulfill the momentum conservation in the transversal plane ($x^{lab}$, $y^{lab}$) within this resolution.

The recoil-ion momentum in direction of the impact $p_{z,rec}$ is not continuously distributed because it is determined by momentum conversation in z direction and by energy conservation. We used the measured $p_{z,rec}$ to determine the final state binding energy of the $He^+$ and to once more reduce the statistical background by testing for energy conservation [20]. In the data analysis we consider only those events were the electron was captured into the 1s state of the projectile ion.

### III Results and Discussion

In order to eliminate the laboratory frame rotation symmetries we present the electron emission pattern in a frame of reference which is defined by the nuclear scattering plane: because electron momentum components in the transversal plane are negligible compared to the nuclear momentum exchange, we simply use the transversal recoil-ion momentum and the direction of impact to define the nuclear scattering plane ($x$, $z$). The projection of the three dimensional electron distributions onto the scattering plane is named "top view". A second type of presentation used in many publications is the so-called "side view" where the electron distribution is projected to the perpendicular plane, which is defined by the vector perpendicular to the scattering plane (y axis) and the direction of impact (z axis). Figure 2(e) illustrates these two planes.

We start the presentation of the results with selected examples of two dimensional electron distributions in planes containing the direction of impact shown in Fig. 2 and Fig. 3. The purpose of these figures is to motivate why the plane perpendicular to the direction of impact need to be investigated to give a complete overview of all our results.

For the highest measured impact velocity $v_p$ = 0.775 a.u. (laboratory frame projectile energy = 15 keV/u) the electron emission pattern mainly consists of two crescent shaped (or "banana" shaped) areas of similar intensity centered in the scattering plane [Fig. 2(a+c)]. These areas extend from the velocity space location of the target nucleus to the projectile nucleus velocity. In the top view presentation shown in Fig. 2(a) the recoil-ion momentum points upwards and the projectile is scattered downwards. Atomic units ($e = m_e = \hbar = 1$) are used and therefore electron velocities and momenta are identical. We use electron velocities because the electron emission pattern is dominantly determined by the location of the nuclei in velocity space.

The velocity of the initial state of the target defines the origin of the coordinate frame. Electron velocities are scaled by the projectile velocity $v_p$. Therefore the projectile is approximately found at (x,y,z) = (0,0,1). The change of both projectile and target nuclear velocities during the collision is below 1 % and not visible on the scale of Fig. 2.

The two-banana structure in Fig. 2a is the signature of a molecular promotion dominated by the $2p\pi_u$ channel. This symmetry of the collision system enforces the dipole lobes of the $\pi$ states to be oriented in the scattering plane. In the side view presentation shown in Fig. 2c the two bananas lie on top of each other which results in a single narrow distribution. Because there is no observable which could be used to define the direction of the y axis the electron distribution has to be symmetric with respect to the scattering plane. We have confirmed that our data are mirror symmetric to with respect to the operation y ➔ -y and then added the counts in both regions in order to reduce the statistical error.

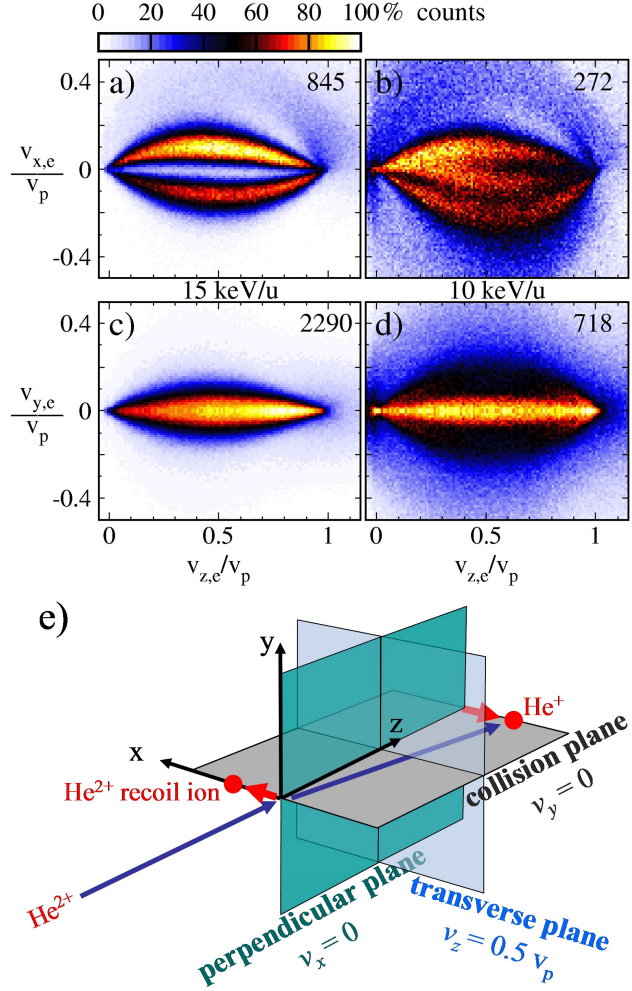

Fig. 2: Electron velocity distribution of the reaction $He^{2+} + He(1s^2) \rightarrow He^+(1s) + He^{2+} + e^-$ in units of the laboratory frame projectile velocity $v_p$. (a), (b) are projections to the nuclear scattering plane (top view). (c) and (d) are projections to the perpendicular plane (side view). The number of measured events per bin (0.01 x 0.01) is given as a linear color scale with the number of counts the maximum number of counts shown at the upper right of each panel. (a) and (c) are measured at 15 keV/u projectile energy and contain only events with small nuclear momentum exchange $p_{r,rec}$ = 1 to 4 a.u. (b) and (d) show the equivalent distributions at 10 keV/u and higher nuclear momentum exchange of $p_{r,rec}$ = 10 to 15 a.u. (e) depicts the geometric definition of the planes.

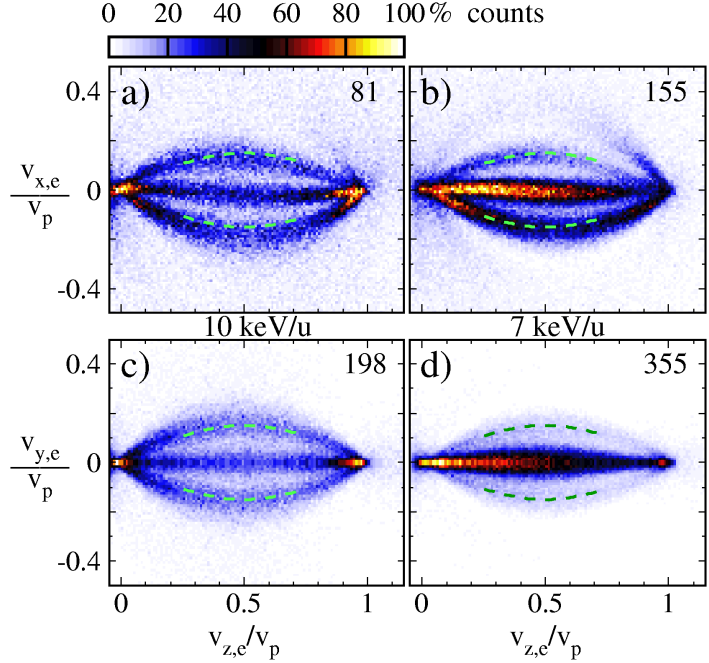

Fig. 3: Slices of the electron velocity distribution within the scattering plane (a,b) and the perpendicular plane (c,d): only events with an out-of-plane velocity smaller than 0.04 $v_p$ are shown. For both impact energies 10 keV (a,c) and 7 keV /u (b,d) a range of transversal recoil momenta $p_{r,rec}$ = 10 to 15 a.u. is selected. The maximum number of counts (i.e. 100% of the color scale) is given at the upper right of each panel. The top and side view presentations corresponding to (a) and (c) are shown in Fig. 2(b+d).

The measured double-banana structure emerges from a long series of quasi-molecular transitions. The first step of this ladder is the promotion to the $2p\pi_u$ orbital. The further evolution of this distribution all the way to the continuum is similar to what one obtains from those classical trajectory calculations which motivated the concept of the saddle point promotion: those electrons which reach the middle plane of the quasi-molecule at small internuclear distance and fail to follow one of the nuclei will be ionized. But as soon as they get out of the middle plane of the quasi-molecule they will be focused to one of the nuclei by the attractive nuclear Coulomb potential. The RLTDSE calculation shows, that with this focusing the number of nodes in the transverse plane is conserved but the original *2pπ_u* distribution is stretched to become the double-banana structure.

By slightly decreasing the impact energy the single differential cross section $d\sigma/dp_{r,rec}$ significantly shifts to a larger nuclear momentum exchange [13]. This effect is much stronger than one might expect from the small increment of the collision time. As a second examples we selected a projectile energy of 10 keV/u and a momentum range $p_{r,rec}$ = 10 to 15 a.u.. The top and side view spectra shown in Fig. 2(b+d) show a structure with more distinct features than the simple double-banana shapes resulting from the 2pπ$_u$ quasi-molecular promotion. To unravel the channels and intermediate quasi-molecular orbitals which give rise to these structures, we switch to a more differential presentation of the data: while in Fig. 2 the data have been integrated over the velocity space dimension which is not shown, we now select a subset of the dataset for which the momentum component out of the plane is very small ($\leq 0.04\ v_p$). Fig. 3(a+c) show such slices. Surprisingly, we find similar structures in both planes consisting of three stripes with the outer stripes shaped like the double-bananas. A further reduction of the projectile impact energy from 10 keV to 7 keV/u affects the structure in the scattering plane only marginally. But in the perpendicular plane the structure contracts to a single line, comparable to the side view structure observed at 15 keV/u.

Fig. 3 shows that the stripes seen in both planes are approximately bend according to the function $C \cdot \sin(\pi\ v_z/v_p)$ which is shown as green dashed lines. We used $C = \pm 0.15$ in all panels of Fig. 3. For 7 keV/u the distributions show a significant asymmetry with respect to the middle plane defined by $v_{z,e}\ /\ v_p = 0.5$. Within the model of quasi-molecular promotions such asymmetries indicate contributions of intermediate states of molecular orbitals such as 3dπ$_g$ [12] which have odd symmetry with respect to the middle plane of the quasi-molecule. For simplification of the further discussions we focus on this middle plane where those states do not contribute. As a technical detail we mention that for analyzing the electron emission pattern in the transverse middle plane [light blue plane if Fig. 2(b)] we have to integrate over a broad range of $v_z$ in order to have sufficient statistics to subdivide our datasets into many regions of nuclear momentum transfer. We included events with $0.25 < v_{z,e}\ /\ v_p < 0.75$. In this region of Fig. 3 we plotted the sine function. At both ends of the integration region the transversal electron distribution (x,y) has shrunken to about 71% of its size in the middle plane. To not obscure the structure by the z integration we scale the x and y components of the electron velocity by $1/\sin(\pi\ v_{z,e}\ /\ v_p)$ before integrating over $0.25 < v_{z,e}\ /\ v_p < 0.75$. The resulting transverse plane distributions of all six measured projectile energies are shown in Figures 4 to 9.

### A. Transverse plane electron emission pattern

Fig. 4 shows the transverse plane electron emission pattern of the measurement at the highest projectile energy of 15 keV/u. The related top and a side view spectra at this energy for small $p_{r,rec}$ are shown in Fig. 2(a+c). The 2pπ$_u$ structure dominating at transversal recoil momenta $p_{r,rec}$ up to 13 a.u shows up as two peaks and resembles a textbook example of a dipole distribution. The dipole axis is vertical, which is the direction of nuclear momentum exchange. The spectra differ from the dipolar shape only at high nuclear momentum transfer. The numbers located at the upper right of each panel correspond to the number of entries of the strongest bin of the spectrum. From these values it can be seen, that this is only a minor contribution to the total cross section. This is why these non $2p\pi_u$ contributions have so far not been resolved in earlier measurements with less resolution and statistics [13]. The two spots at positive and negative $v_{x,e}$ do not have exactly the same intensity. This results from an additional contribution of an $1s\sigma_g$ quasi-molecular promotion which adds up coherently to the $2p\pi_u$ peaks and therefore increases one of the spots while the other's intensity is reduced [21].

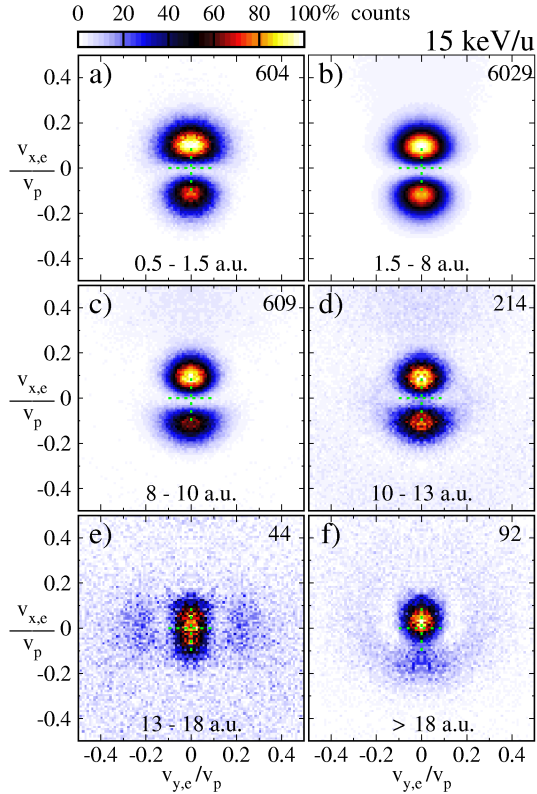

Fig. 4: Scaled electron velocity distribution within the plane perpendicular to the direction of impact for an impact energy of 15 keV/u (projectile velocity $v_p$= 0.77 a.u.) for different regions of recoil momentum transfer $p_{r,rec}$ as stated at the bottom of each panel. The measured velocities have been divided by $\sin(\pi\, v_{z,e} / v_p)$ before integrating in beam direction from $v_{z,e}$ = 0.25 $v_p$ to 0.75 $v_p$. The transversal momentum transfer to the recoil-ion defines the x-axis, therefore the projectile is scattered downwards. The number of measured events is represented by a linear color scale with the maximum number of counts given at the upper right of each panel. The data are mirrored with respect to the scattering plane ($v_{y,e}$ =0) in order to reduce the statistical error. For nuclear momentum transfers above 12 a.u. the recoil-ion spectrometer does not have full solid angle of detection.

Next we show a representative sample of recoil-ion transversal momentum regions $p_{r,rec}$ from our comprehensive dataset. A more detailed presentation of the data can be found in the supplementary material. By visual inspection of that data set we have selected regions of $p_{r,rec}$ in which the electron emission pattern is rather constant and then integrated over $p_{r,rec}$ to improve the statistics. This yielded typical and representative patterns shown in Figures 4-9.

The contributions of quasi-molecular promotion beyond $1s\sigma_g$ and $2p\pi_u$ become more visible when reducing the projectile energy. Fig. 5 shows the results recorded at 13.3 keV/u, Fig. 6 those obtained at 11.5 keV/u and Fig. 7 those for 10 keV/u projectile energy. For the case of the highest internuclear momentum exchange investigated we find a ring structure with a sharp spot at the centre. This structure is at the center of our spectrum which is defined by the internuclear axis of the final state. Nevertheless we believe that it is caused by a quasi-molecular promotion starting with the $2s\sigma_g$ state. The shift to positive values of x and the higher intensity at negative x which is seen in Fig. 5(f), Fig. 6(h) and Fig. 7(l) can be explained by a $2p\pi_u$ contribution, which is coherently added to the $2s\sigma_g$ channel.

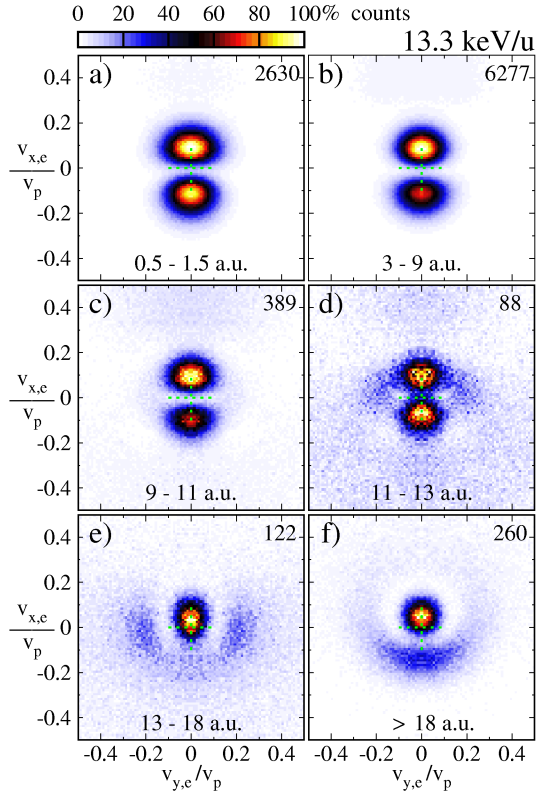

Fig. 5: Scaled electron transversal velocity distributions similar to Fig. 4 but for a smaller impact energy of 13.3 keV/u ($v_p$= 0.73 a.u.)

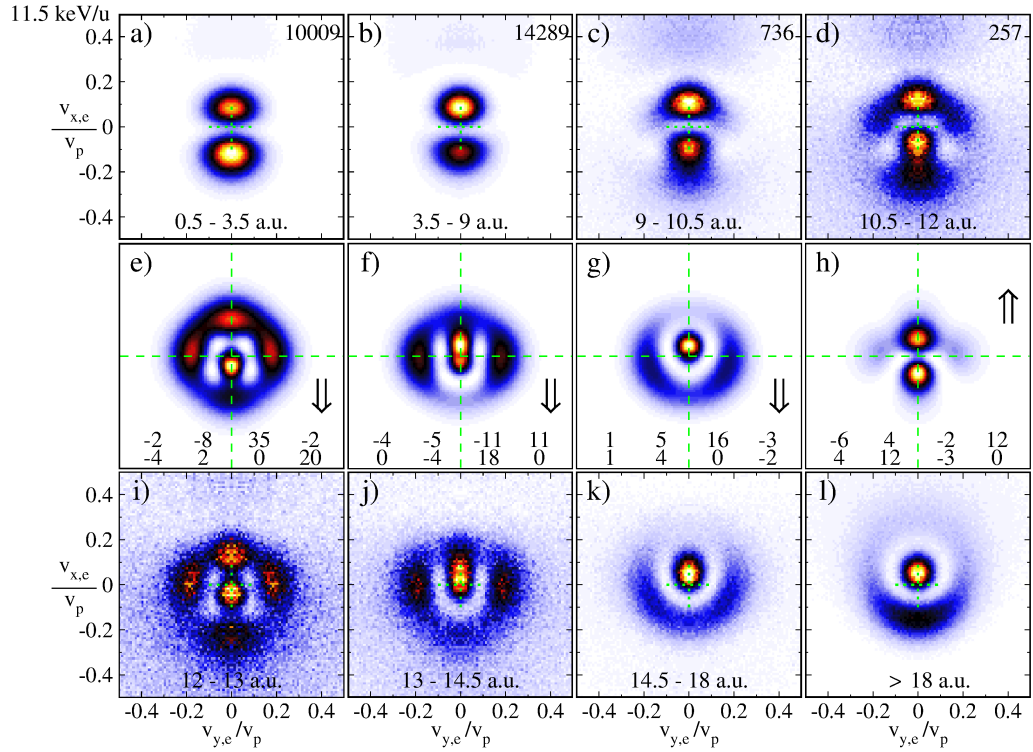

Fig. 6: Electron transversal velocity distributions. (a-d) (i-l) Scaled experimental distributions recorded at a projectile energy of 11.5 keV/u (projectile velocity $v_p$= 0.68 a.u.). (e-h) Distributions modeled by the four states pictured in Fig 10(a-d). The real and imaginary part (upper and lower value, respectively) of the four coefficients of the $1s\sigma_g$, $2p\pi_u$, $2s\sigma_g$ and $3d\delta_g$ contributions are given in each panel.

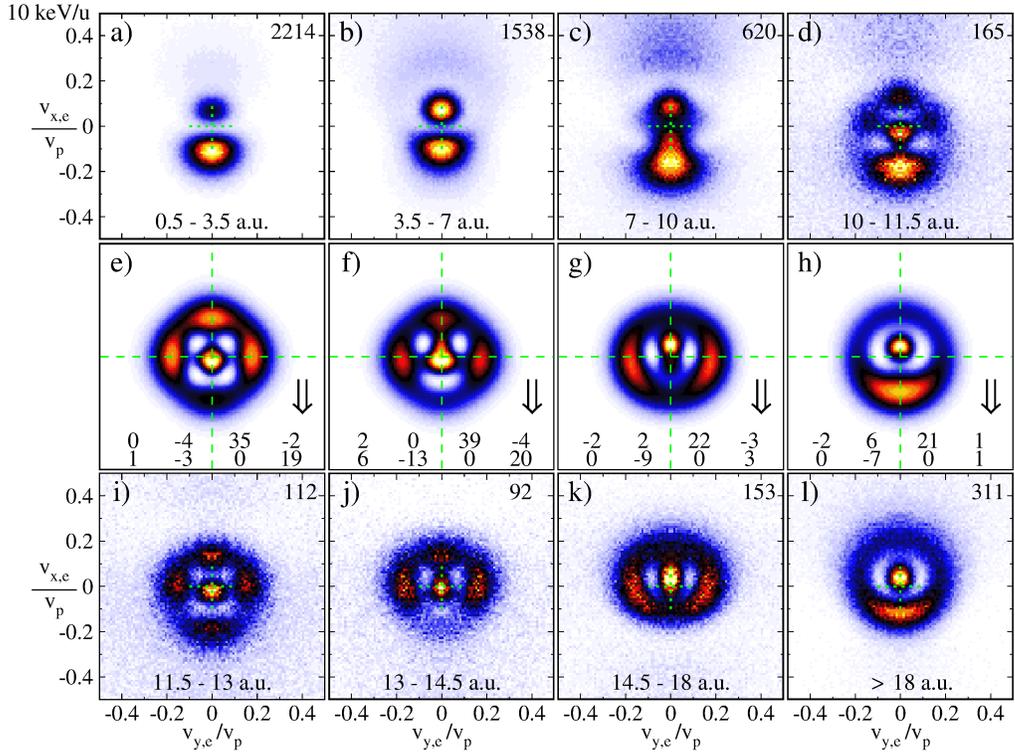

Fig. 7: Electron transversal velocity distributions. (a-d) (i-l) Scaled experimental distributions measured at a projectile energy of 10 keV/u (projectile velocity $v_p$= 0.63 a.u.). (e-h) Distributions modeled by the four states pictured in Fig. 10(a-d). The real and imaginary part (upper and lower values, respectively) of the four coefficients of the $1s\sigma_g$, $2p\pi_u$, $2s\sigma_g$ and $3d\delta_g$ states are given in each panel.

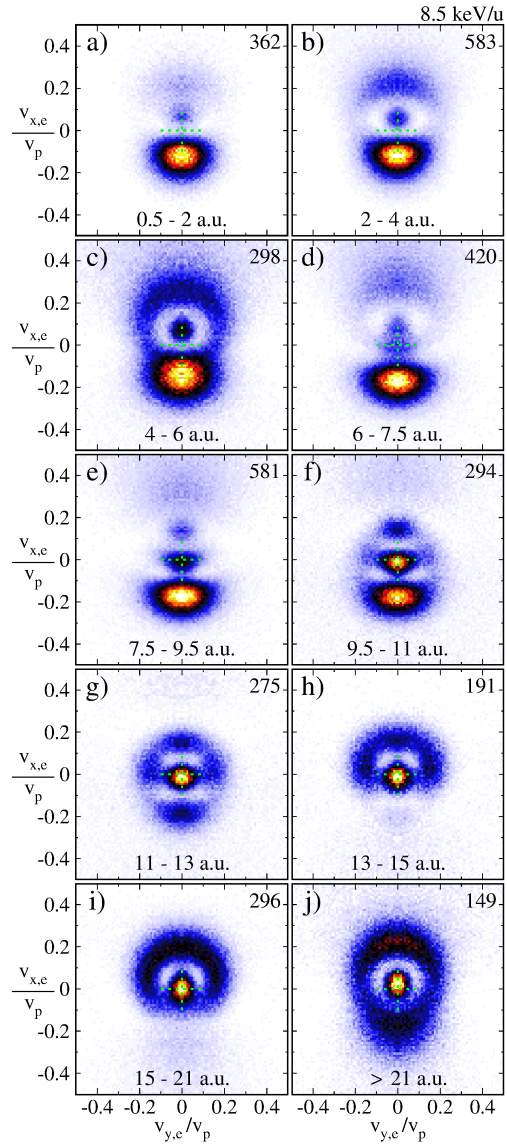

Fig. 8: Scaled electron transversal velocity distribution similar to Fig. 4 but for a smaller impact energy of 8.5 keV/u (projectile velocity $v_p$= 0.58 a.u.)

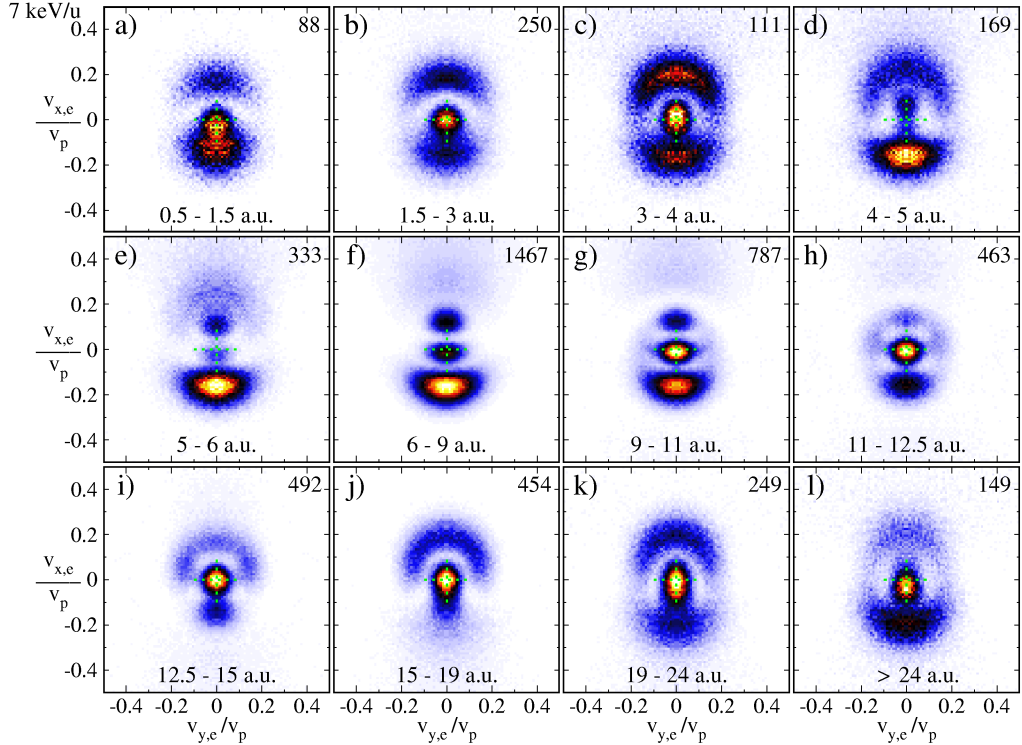

Fig 9: Scaled electron transversal velocity distribution similar to Fig. 4 but for 7 keV/u projectile energy ($v_p$= 0.53 a.u.)

### B. Modeling of the experimental data by four channels of quasi-molecular promotion

To the present day, a sophisticated theoretical description is only available for the emission pattern depicted in Fig. 7(i) which shows events with $p_{r,rec}$ of approx. 12 a.u. for a projectile energy of 10 keV/u. As discussed in detail in [12] the structure observed is caused by the superposition of the $2s\sigma_g$ and $3d\delta_g$ channel. If the relative phase between the wave functions of these two channels is such that the cross term vanishes, then the four lobes of the $3d\delta_g$ contribution fill up the ring-shaped node of the $2s\sigma_g$ contribution. If the wave functions of these two contributions are chosen real valued a relative phase of 90° results in a vanishing of the cross term.

Because the collision system is symmetric with respect to the scattering plane two of the lobes have to be centered in the scattering plane which is plotted vertically with the recoil scattered upwards. The difference between the upper and the lower part is most probably caused by a small $2p\sigma_u$ contribution as mentioned above.

Similar electron emission patterns which are also dominated by the superposition of $3d\delta_g$ and $2s\sigma_g$ channel can be found at 11.5 keV/u projectile energy with $p_{r,rec}$ from 12 to 13 a.u. [Fig. 6(i)] and 8.5 keV/u with 11 to 13 a.u. [Fig. 8(g)]. Even at projectile energies of 7 keV/u with $p_{r,rec}$ between 11 a.u. and 12.5 a.u. this structure can be identified [Fig. 9(h)], here however the lower peak is strongly enhanced.

In the following section we will investigate how much of the structures seen at the six projectile energies can be explained by restricting to the four quasi-molecular channels we already identified and which have been described in [12]. Therefore we examine what variety of electron distributions $P(v_{e,x},v_{e,y})$ can be obtained by superimposing only four 2D model wave functions using complex coefficients $c$.

$$P(v_{e,x},v_{e,y}) = |\ c_{1s\sigma}\ \psi_{1s\sigma} + c_{2p\pi}\ \psi_{2p\pi} + c_{2s\sigma}\ \psi_{2s\sigma} + c_{3d\delta}\ \psi_{3d\delta}\ |^2$$

The model wave functions are constructed as $\psi(v_r,\varphi) = R(v_r)\cdot\cos(m\ \varphi)$. The azimuthal angle $\varphi$ is related to Cartesian coordinates by $\varphi = \mathrm{atan}(v_{y,e}/v_{x,e})$ and $v_{r,e}^2 = v_{x,e}^2 + v_{y,e}^2$. The quantum number m = 0, 1, 2 describes the σ π and δ state. An accurate radial part of the velocity space wave function $R(v_r)$ could be calculated by the RLTDSE method but for the present work we simply use model functions that fit best to the data. In Fig. 10(a-d) we present the 2D densities of these model wave functions. The areas of high densities are labeled with the sign of the wave function which is real-valued.

For illustration Figs. 10(e+f) show the basic structures arising from only two of the model wave functions. As mentioned above a slightly asymmetric dipole distribution as occurring in (e) results from $2p\pi_u$ and $1s\sigma_g$ contributions added with a relative phase of 0 or 180°. A phase difference of 90° would not lead to asymmetric peaks but cause a vanishing of the horizontal nodal line.

A superposition of $2s\sigma_u$ and $3d\delta_g$ channels with approximately the identical magnitude and a phase difference of 90° is shown in Fig. 11(f). A remarkable aspect of this structure is the phase evolution close to the four local minima: on a circular path around each of these nodes the phase of the wave function changes by 360°. As described in detail in [12] this phase development is related to a quantum mechanical flux and therefore the experimental observation of this structure supports the theoretically predicted vortices occurring in the wave function of a single free electron.

Several of the experimental distributions can be modeled in surprising detail by superimposing the four quasi-molecular channels. Figs. 6(e-h) and 7(e-h) show a few examples of the modeled distributions. The arrows drawn inside these spectra point towards the corresponding experimental distributions. In each panel of these figures the real parts of the four coefficients are given in the upper line and the imaginary part in the lower line of values. The most left column belongs to the coefficient of the $1s\sigma_g$ contribution followed by $2p\pi_u$, $2s\sigma_g$ and $3d\delta_g$ coefficients. We adjusted the coefficients manually to best reproduce the measured data. Because the absolute phase of the model wave function does not affect the density and the simulated distributions are not normalized it would be possible to reduce the needed number of parameters to 6. Employing a numerical fitting procedure yielded much less satisfactory results - even with well adjusted start values of the fit parameters. This is probably due to the poor modeling of the radial part of the distribution.

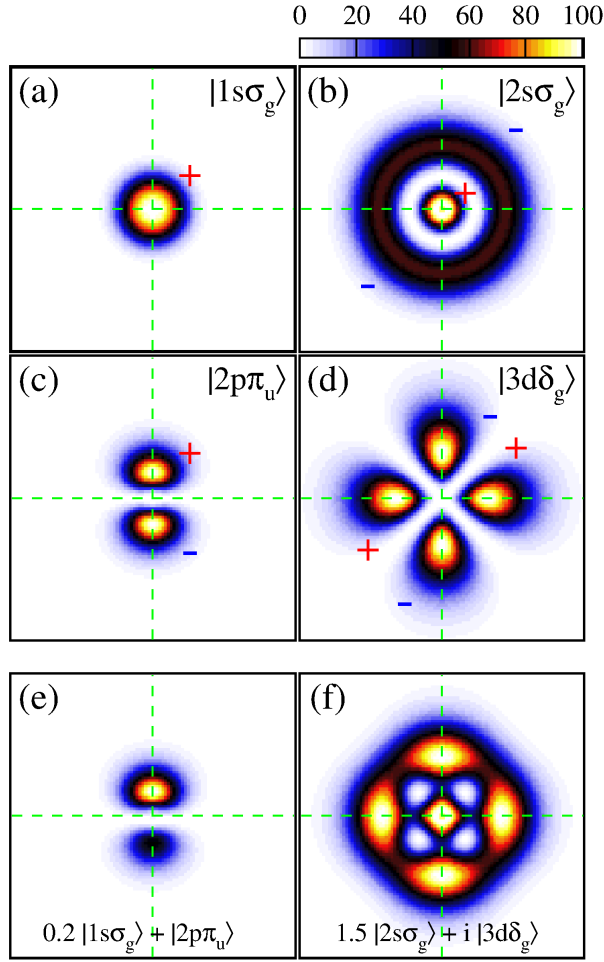

Fig. 10: Density of the $1s\sigma_g$, $2p\pi_u$, $2p\sigma_g$ and $3d\delta_g$ contributions in the transverse plane which are used to model the measured spectra. The related wave functions are real-valued. The radial component was adjusted by hand to best match the measured contributions. (e) and (f) show typical distributions which can be obtained by superimposing only two of these contributions.

The major differences between the modeled and experimental distributions are contributions at the edges of the spectra (high $v_{r,e} = \sqrt{v_{x,e}^2 + v_{y,e}^2}$). These contributions are isotropic in many regimes of collision energy and nuclear momentum transfer. However at internuclear momentum transfers between 3 a.u. and 10 a.u. we see a small contribution directed towards the recoil side [upper part of the spectra; see for example Figs. 4(b), 5(b), 6(b) 7(c), 8(d) and 9(f)]. Similar contributions at the bottom of the spectra which corresponds to the direction of the projectile transversal momentum appear at higher internuclear momentum transfers. Therefore, these can only be seen at the lower impact energies [see for example Fig. 8i) and Fig. 9(j)].

To visually enhance these contributions we present the corresponding subsets of the measurements at projectile energies of 7 keV/u and 8.5keV/u as polar plots [Fig 11(a-d)]. These one dimensional distributions can now be fitted using our model function for fixed $v_{r,e}$ employing standard fitting algorithms. We have restricted our fit to either only wave functions of σ, π, and δ symmetry (m ≤ 2, blue lines) or we used φ-states in addition (m ≤ 3, read lines). Details of the fitting procedure are described below. Over 500 Fits of this type have been performed for all projectile energies, for small regions of transversal recoil momentum $p_{r,rec}$ and for small regions of transversal electron velocity with a width of 0.02 $v_p$. Figs. 11(e+f) show only two of these results in regimes where the δ symmetry dominates the angular distribution. The main result of the fitting is, that within the experimental uncertainty we did not find any contribution of angular momenta higher than m = 2. The fitting results of m ≤ 3 and m ≤ 2 typically differ less than the statistical error of the measured data and the lines representing the fitting results can hardly be separated. Therefore they are plotted separately on the left and the right of the polar plots in Fig. 11. As mentioned above the experimental data have been mirrored in order to reduce the statistical errors.

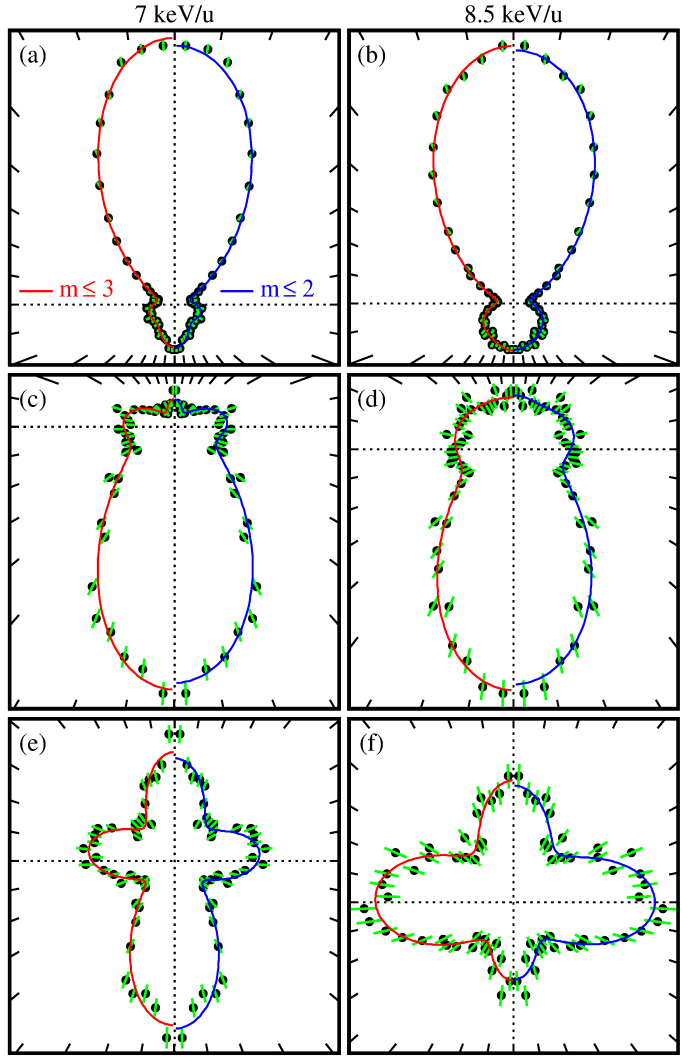

Fig. 11: Electron angular distributions in the transverse plane for 7 keV/u (a, c, d) and 8.5 keV/u (b, d, e) projectile energy in a polar representation. The momentum transfer to the recoiling ion points upwards. The experimental results (cycles with green error bars) contain events with (a+b): 6 a.u. < $p_{r,rec}$ < 8 a.u., 0.4 $v_p$ < $v_{r,e}$ < 0.47 $v_p$ [compare to Fig. 9(f) and 8(d)]. (c+d): 15 a.u. < $p_{r,rec}$ < 19 a.u., 0.4 $v_p$ < $v_{r,e}$ < 0.47 $v_p$ [compare to Fig. 9(j) and 8(i)]. (d+e): 11 a.u. < $p_{r,rec}$ < 12.5 a.u., 0.1 $v_p$ < $v_{r,e}$ < 0.12 $v_p$ [compare to Fig. 9(h) and 8(g)]. The data have been fitted under the assumption that quasi-molecular channels with angular momentum up to m = 3 (red line, plotted at the left side of each plot) or only up to m = 2 (blue line, at the right) contribute (see Eq. 1).

## C. Mechanism of electron promotion to the continuum

Before discussing the fitting results in detail we would like to speculate on the process which produces the outer loops on either the recoil or the projectile side. We will discuss this process within a single active electron model. using $H_2^+$ states. The initial state with the electron located at the target nucleus is described by a superposition of $1s\sigma_g$ and $2p\sigma_u$ states which are sketched in Fig. 12(a+d). During the approach of the nuclei on the incoming part of the collision trajectory $3d\sigma_g$ and $2p\sigma_u$ states can be populated by radial coupling. These couplings start to occur at internuclear distances of about 10 a.u. and reach a maximum coupling strength at about 5 a.u. [22]. Thus the $3d\sigma_g$ and $2p\sigma_u$ states get populated early enough during the collision, so that these distributions are able to rotate into the transverse plane during later collision times as sketched in Fig. 12(c+f). These rotational couplings populate does not populate a pure quasi-molecular state but a coherent superposition of several molecular states. However, most simply one can expect the electric wave function (with some probability) to remain frozen in space while the nucleus passes by which a rapid rotation of internuclear axis is caused. We notice that it is not trivial to map this simple picture based on single electron quasi-molecular states to the correlation diagram of $He_2^{2+}$.

If the impact parameter is not too small the gerade and the ungerade channel (upper and lower part of Fig. 12) will stay in phase and therefore the electron is finally expected to be found at the recoil side where it was initially located. This was experimentally demonstrated in Fig. 11(a+b) where an intermediate nuclear momentum exchange was selected. Close collisions result in higher recoil transversal momenta and induce a phase shift between gerade and ungerade states because of their different binding energies at small internuclear distances. Therefore the lobes which constructively added up move to the projectile side as shown in Fig. 12(c+d).

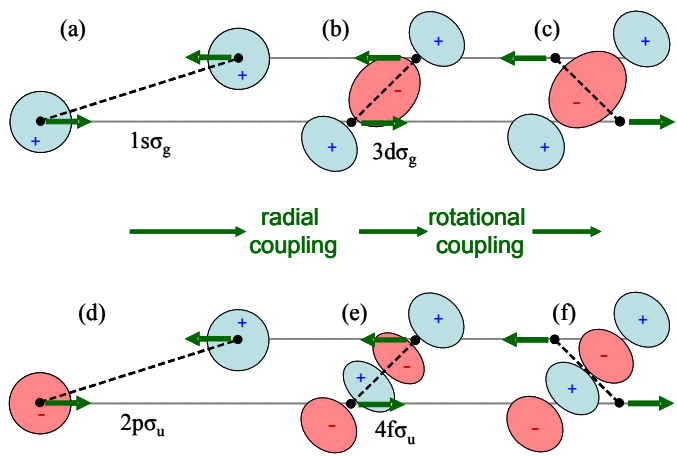

Fig. 12: Illustration of a the population of high angular momentum quasi-molecular states within a single active electron model. (a) - (c) show the case of an initial $1s\sigma_g$ quasi molecular orbital, (d) - (f) the corresponding process occurring for $2p\sigma_u$ states.

## D. Amplitudes and phases of σ, π and δ states as a function of the radial electron momentum

Despite the success in modeling some of the electron distributions with only 4 basis states it is obvious that more quasi-molecular states are needed. As mentioned above we can assume that only states of σ, π and δ symmetry significantly contribute. To obtain the radial wave function of a specific angular momentum component we fit the electron angular distributions in small regions of electron transversal velocities $v_{r,e}$ and recoil transversal momentum transfer $p_{r,rec}$ using only the angular parts of the 2d-model functions shown in Fig. 10.

For $\varphi = 0$ the electron is emitted in direction of the recoil transversal momentum (upwards in Fig. 3 to Fig. 11) and $\varphi = \pm 90°$ is perpendicular to the nuclear scattering plane (i.e. horizontal in these figures). Only wave function with an angular component of $\cos(m\varphi)$ contribute because the nuclear dynamics which defines the potential of the electronic problem is symmetric with respect to the collision plane. We start our fitting procedure with angular momenta up to $m = 2$ and a complex coefficient $g_m$. The fitted electron angular distribution is given by

$$P^m(\phi) = \left| \sum_{m=0,1,2} g_m \cos(m\varphi) \right|^2 \quad (1a)$$

$$= [a + b\cos(\varphi) + c\cos(2\varphi)]^2 + [d + e\cos(\varphi)]^2 \quad (1b)$$

Because the absolute phase of the wave function does not effect the measurable distribution we can restrict one of the coefficients $g_m$ to a real values. We selected $g_2$ which belongs to the δ contribution to be real valued.

For further discussion it is helpful to transform Eq. 1a into an equation with real valued parameters only. The result is given as Eq. 1b where the parameters $a$, $b$ and $c$ are the real parts of $g_0$, $g_1$ and $g_2$ and $d$ an $e$ are the imaginary parts of $g_0$ and $g_1$. We employed the program library MINUIT [23] (developed at CERN) to obtain values of these parameters which yield the best agreement between $P(\phi)$ and our measured distributions.

This fitting does not provide a unique solution. If the signs of $a$, $b$ and $c$ are switched $P(\phi)$ does not change. The same problem appears with $d$ and $e$. In addition to these trivial ambiguities of the phases it is possible to achieve different results for the absolute values of the σ, π and δ contribution. Furthermore, the identity $2\cos^2(\varphi) = 1 + \cos(2\varphi)$ allows us to construe the probability distribution of a pure π state as the sum of an isotropic σ contribution and the cross term between σ and δ.

The ambiguities in the fitting procedure can be avoided by using orthogonal basis functions to fit the measured electron angular distribution:

$$P^n(\phi) = \sum_{n=0,1,2,3,4} k_n \cos(n\varphi) \qquad (2)$$

The coefficients $k_n$ have no immediate physical meaning but they are linked to the coefficients of Eq. 1b by

$$k_0 = a^2 + d^2 + (b^2 + c^2 + e^2)/2 \qquad (3a)$$
$$k_1 = 2(ab + de) + bc \qquad (3b)$$
$$k_2 = (b^2 + e^2)/2 + 2ac \qquad (3c)$$
$$k_3 = bc \qquad (3d)$$
$$k_4 = c^2 \qquad (3e)$$

Not all distributions represented by $P^n$ can be generated by a fitting using Eq. 1. Obviously the weighting factors $k_4$ and $k_0$ have to be positive to get an angular distribution of the type of $P^m$ which is restricted by the assumption that only states with even symmetry with respect to the scattering plane contribute.

We analyzed how many different sets of coefficients $a$, $b$, $c$, $d$ and $e$ lead to the same $k_m$. Without loss of generality we set $c = \sqrt{k_4}$ which eliminates one ambiguity as previously mentioned. Thereby $b$ can easily be calculated by using Eq. 3d. In case of $c \neq 0$ the Eqs. 3a to 3c can be merged to a cubic equation which determines the coefficient $a$. This equation provides up to three solutions but not more than two of these solution lead to complete set of real valued coefficients $a$, $b$, $c$, $d$ and $e$.

Figures 13(a-c) and Fig. 14 show the absolute values of the coefficients $|g_0| = (a^2 + d^2)^{0.5}$ (black lines) $|g_1| = (b^2 + e^2)^{0.5}$ (read lines) and $|g_2|=c$ (green line). The sizes of the rectangles represent the uncertainties given by the MINUIT library. In most cases we obtained two results of contradictory meaning which can be distinguished by the absolute value of the coefficient $g_1$ ($\pi$ contribution). This enables us to connect the two equivalent fitting results for each $v_{r,e}$ to a radial function which is the absolute value of the radial parts of a wave function. The absolute value of the $\delta$ amplitude $c = \sqrt{k_4}$ is the same for both solutions because $k_4$ is a weighting factor of an orthogonal set of base functions. In case of small $\delta$ contribution the ambiguity of the $\sigma$ and $\pi$ contributions disappear as seen for example at the higher $v_{e,r}$ in Fig. 13(a).

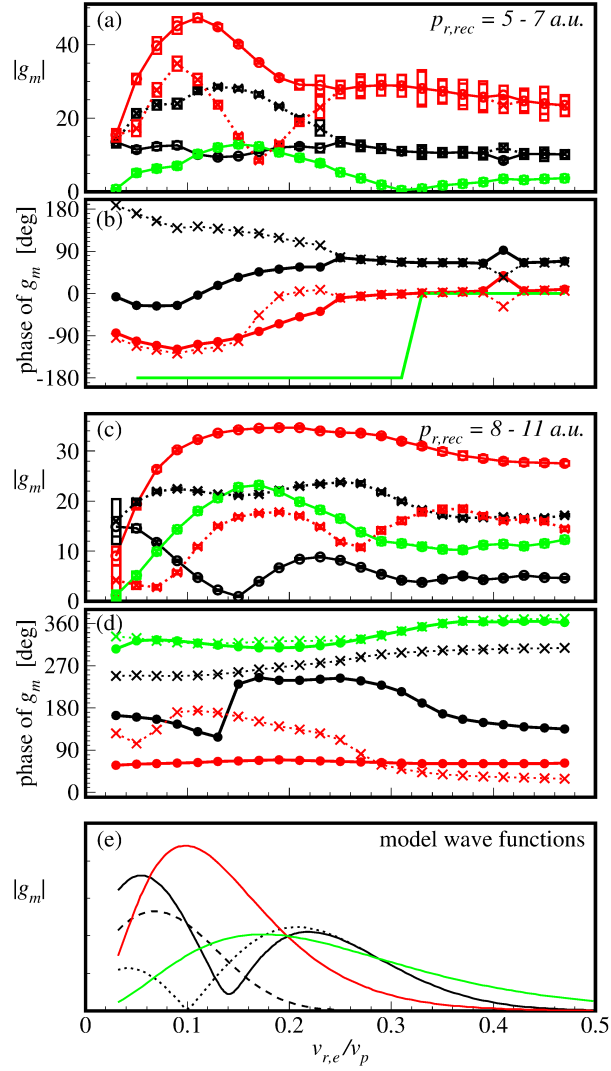

Fig. 13: Amplitudes of the σ (black, m = 0), π (red, m = 1) and δ (green, m = 2) contribution as a function electron transversal velocity $v_{e,r} = (v_{e,x}^2 + v_{e,y}^2)^{0.5}$ at 10 keV/u projectile energy. The experimentally determined absolute values $|g_m|$ of the coefficients are shown for recoil transversal momenta $p_{r,rec}$ = 5 - 7 a.u. (a) and $p_{r,rec}$ = 8 - 11 a.u. (b). In most cases the fitting procedure provides two sets of $g_m$ leading to identical angular distributions. The related phases of the $g_m$ are shown in (b) and (d). The indeterminacy of the phases was reduced by minimizing the change of the coefficients with $v_{e,r}$. At $p_{r,rec}$ = 5-7 a.u. (b) the δ contribution $g_2$ is restricted to real values. (e) radial component of the wave functions visualized in Fig. 9(a-d). Three σ states are shown as black lines: 1sσ (dashed line) 2sσ (dotted line) and a superposition of 1sσ and 2sσ with 30° relative phase (solid line)

As explained above, the phases of the coefficients $g_m$ can not explicitly be determined. But it is possible to arbitrarily select one of the possibilities at a specific $v_{e,r}$ and to connect the phases at the other values of $v_{r,e}$ under the assumption that the changes of $g_m$ are minimal. Figs. 13(a+b) show the resulting curves for 10 keV/u projectile energy and recoil transversal momenta $p_{r,rec}$=5-7 a.u. Here $g_2$ is still real valued but to avoid artificially strong phase changes of the σ and π contribution we have to allow negative values of $c = g_2$ which causes a phase jump of 180° at $v_{r,e}$ = 0.3 $v_p$ when the sign changes.

Obviously the phase related to the two possible absolute values of $g_0$ and $g_1$ have to differ to yield the same angular distributions. This is due to the cross terms between the σ, π and δ distributions. The line types in Fig. 13 indicate which phase evolutions belong to which absolute values of the coefficients.

The red dotted curve of the π contribution $g_1$ in Fig. 13a shows a local minimum at $v_{r,e}$ = 0.17 $v_p$ without going to zero. Such a structure is produced when, for example, $b$=Re($g_1$) changes from negative to positive values and $d$=Im($g_1$) stays constant. We can assume that two HC-channels of π symmetry add up with a relative phase of approx. 90° and one of the radial components has a node.

Fig. 13(c) shows the results at $p_{r,rec}$ = 8 to 11 a.u. Here the δ contribution shows only a very weak local minimum at the location where $c$ showed a zero crossing in Fig 13(a+b). Therefore we conclude that even the δ contribution consists of more than one HC-cannel with different radial components which add up with a relative phase of about 90°. Thus it is not appropriate to restrict $g_2$ to real values. When preprocessing the fitting results obtained at different $v_{r,e}$ to connect them to the curves shown in Fig. 13(d) we minimized the change of all three coefficients $g_0$, $g_1$ and $g_2$

The zero crossing of $g_0$ shown by the black solid lines in Fig. 13(c+d) can be assigned to the node of the radial wave function of the 2s$\sigma_g$ state. Fig. 14 shoes the absolute value of the coefficients for the same nuclear momentum exchange but different projectile velocities. The position of the minimum of $g_0$ is not fixed but it changes with increasing projectile energy to larger $v_{r,e}$. However, this does not suggest that the 2sσ contribution itself changes in that way. By adding the 1s$\sigma_g$ state coherently to the 2s$\sigma_g$ contribution the minimum of $|g_0|$ can be moved to a $v_{r,e}$ that is different to the position of the node of the 2sσ state. Fig. 13(e) visualizes the radial components of the two-dimensional model functions shown in Fig. 9(a+e). The 1s$\sigma_g$ and 2s$\sigma_g$ parts are shown as dotted and dashed black lines. The solid black line shows the σ amplitude of a superposition of these states. The minimum is slightly filled because 1s$\sigma_g$ and 2s$\sigma_g$ have been added with a relative phase of 30°.

At the lowest projectile energies measured $|g_0|$ has a second local minimum at $v_{r,e} \approx$ 0.3 $v_p$ which is clearly visible in the fitting solution depicted by the solid lines but also present in the other solutions. A σ contribution with a second radial node at slow impact velocities suggests that not only $1s\sigma_g$ and $2s\sigma_g$ but also that the $3s\sigma_g$ HC-channel is relevant.

Fig. 14 shows that the contribution of δ states strongly increases when the projectile energy is reduced from 15 keV/u to 7 keV/u. Contrarily, a corresponding increased production of He(3d)$^+$ by single electron transfer is not predicted by close coupling calculations by Frisch [24] at 20, 12.5, 8, 5.5 keV/u and other energies. The total cross section of the single electron transfer into 3d states shows a maximum at 12.5 keV/u. The electron capture into the 1s state of the projectile with simultaneously excitation of the electron remaining at the target into the 3d state continuously increases from 20 keV/u to 5.5 keV/u but only by about 60 %. At 12.5 keV/u and 8 keV/u this calculation gives a cross section for the production of He$^+$ with $l = 3$ by electron transfer which is more than 5 times lower than the cross section of the $l = 2$ case. The magnetic quantum numbers of the states are not given but one can assume that the ratio of the total cross sections of m=3 and m=2 is much higher than the ratio between d and f states. This is consistent with our finding that m = 4 does not significantly contribute to the transfer ionization.

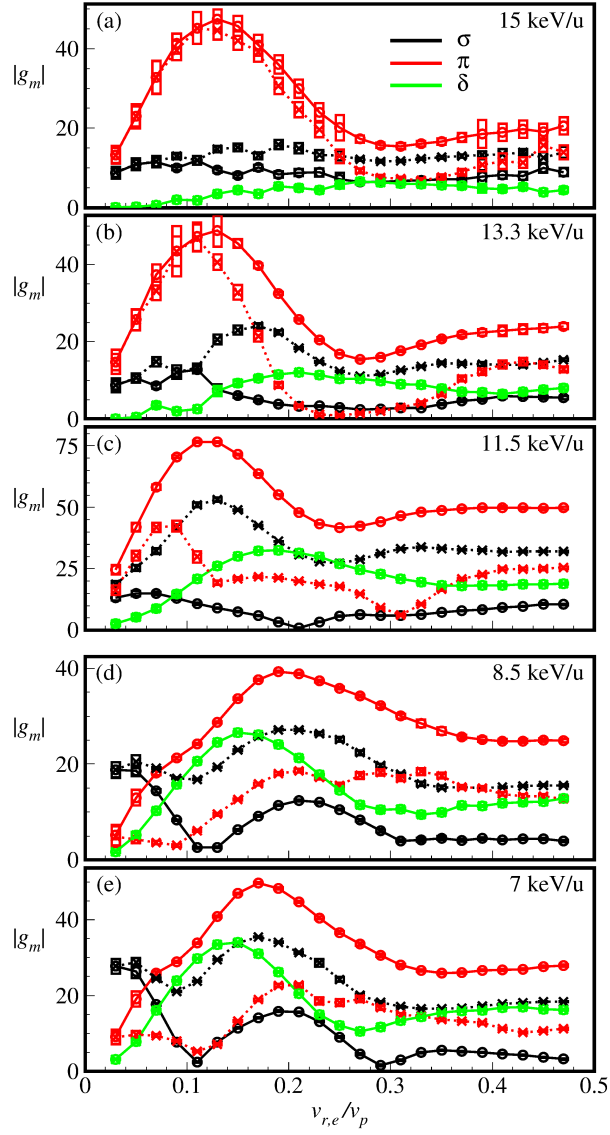

Fig 14: Absolute values of the amplitudes of the σ (black), π (red) and δ (green) contribution as a function of the electron transversal velocity $v_{e,r} = (v_{e,x}^2 + v_{e,y}^2)^{0.5}$ at recoil-ion momenta $p_{r,rec}$ = 8 to 11 a.u. for different projectile energies as labeled [the result at 10 keV/u are shown in Fig. 11(c)]

## VII. Summary

Achieving better statistics and resolution than in previous published experiments enabled us to discover the existence many more HC-channels than only $1s\sigma_g$ and $2p\pi_u$ which have been known since the middle 90[th] and $2s\sigma_g$ and $3d\delta_g$ which have recently been employed by the 2eHC-RLTDSE method [12]. The importance of these additional channels rapidly increases at projectile energies below 10 keV ($v_p = 0.63$ a.u.). The fitting of the electron distributions at the transverse middle plane of the quasi-molecule does not support the existence of higher angular momentum components than $m = 2$ but shows nodes at in the radial component of the wave function. Our fitting of amplitudes and phases, even though not completely unambiguous, will provide a benchmark test for future calculations which should including much more HC channels than the calculations presently available.

The experimental investigations using reaction microscopes easily provide the resolution to separate the electron transfer channels into different shells but only a few experiments provided sufficient resolution so measure the final state angular momentum (e.g. [25]). Therefore the measurement of the electron emission via the saddle point process is a powerful alternative method to experimentally study the angular momentum transfer from the nuclear motion into the electronic system in slow ion atom collisions.

Depending on the nuclear momentum exchange an occurrence of electrons with high radial velocity at either the recoil-ion side or the projectile side was observed at all projectile velocities. This points to a mechanism which leads to fixed relative phases between the contributions of different angular momenta. We suggest that this can be explained by a radial coupling to higher quasi-molecular states of σ symmetry while the projectile approaches followed by rotational coupling.

This work is supported by the Deutsche Forschungsgemeinschaft (SCHM 2382/2-1).

# Supplementary material for
"Quasi-molecular electron promotion beyond the 1sσ and 2pπ channels in slow collisions of He$^{2+}$ and He"

L. Ph. H. Schmidt, M. Schöffler, C. Goihl, T. Jahnke, H. Schmidt-Böcking, R. Dörner

*Institut für Kernphysik, Goethe-Universität, 60438 Frankfurt am Main, Germany*

Using COLd Target Recoil Ion Momentum Spectroscopy (COLTRIMS) we investigated the transfer ionization in slow $^4$He$^{2+}$ + $^4$He collisions:

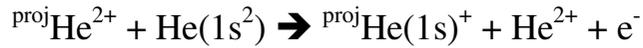

$$^{proj}He^{2+} + He(1s^2) \rightarrow {}^{proj}He(1s)^+ + He^{2+} + e^-$$

At six projectile energies between 7 keV/u (projectile velocity $v_p$=0.53 a.u.) and 15 keV/u ($v_p$=0.77 a.u.) the fully differential cross sections were measured (a.u. denotes atomic units: $m_e$ = e = ℏ = c/137 = 1). Here we show a comprehensive dataset of these results. Momenta and velocities are given in atomic units.

We present the electron emission pattern in a frame of reference, which is defined by the nuclear scattering plane. The measured electron velocities are given in units of the projectile velocity $v_p$. Because the electron momentum components in the transversal plane are negligible compared to the nuclear momentum exchange, we simply use the transversal recoil-ion momentum $p_{r,rec}$ and the direction of impact to define the nuclear scattering plane $(x, z)$. Figure 1(a) illustrates this definition.

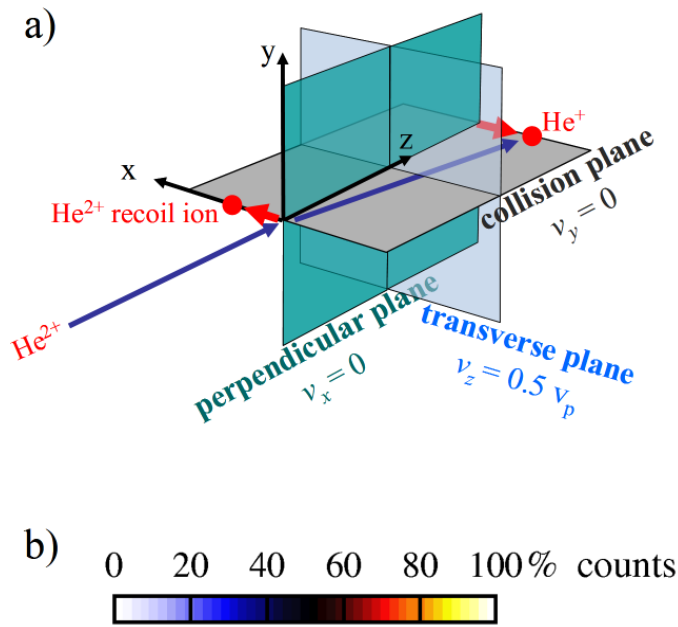

Fig. 1: (a) Definition of the frame of reference defined by the nuclear momenta, which is used to present the emitted electron distributions. The direction of impact defines the z-axis. By definition the y-component of the recoil-ion momentum is zero. (b) Color scale used in Figures 2 to 31.

For each projectile energy and for small regions of nuclear momentum transfer five kinds of two dimensional electron distributions are shown to describe the three dimensional distribution:

**Figures 2 to 7** show the so-called top view presentations, which are projections on the nuclear scattering plane. The projectile, coming from the left side, is scattered downwards. The spectra contain events with out of plain electron velocities up to $0.5v_p$ (integration of the three dimensional distribution from $v_{y,e}$ = -0.5$v_p$ to 0.5$v_p$).

**Figures 8 to 13** show the so-called side view presentations, which are projections on the perpendicular plane. This plane is defined by the direction of impact (z-axis) and the direction perpendicular to the scattering plane (y-axis). The spectra contain events with out of plain electron velocities up to $0.5v_p$ (integration from $v_{x,e}$ = -0.5$v_p$ to 0.5$v_p$). The data are mirrored with respect to the scattering plane ($v_{y,e}$=0) in order to reduce the statistical error.

**Figures 14 to 19** show the scattering plane distributions of the emitted electron. The spectra contain events with out of plain electron velocities up to $0.04v_p$ ($0.04v_p \leq v_{y,e} \leq 0.04v_p$)

**Figures 20 to 25** show the perpendicular plane distributions of the emitted electron. The spectra contain events with out of plain electron velocities up to $0.04v_p$ ($0.04v_p \leq v_{x,e} \leq 0.04v_p$). The data are mirrored with respect to $v_{y,e}$= 0 in order to reduce the statistical error.

**Figures 26 to 31** show the transverse plane distributions of the emitted electron. The measured velocities have been divided by $\sin(\pi \ v_{z,e}\ /v_p)$ before integrating the three dimensional distribution in beam direction from $v_{z,e}$ = 0.25$v_p$ to 0.75$v_p$ .The data are mirrored with respect to $v_{y,e}$=0 in order to reduce the statistical error.

In all panels of all Figures 2 to 31 the number of measured events is represented by a linear color scale with the maximum number of counts given at the upper right of each panel. The regions of recoil momentum transfer $p_{r,rec}$ are stated at the bottom of each panel. From $p_{r,rec}$ = 0.5 a.u. to 15 a.u. we selected regions of 0.5 a.u. width followed by regions of 1 a.u. width up to 20 a.u. and regions of 2 a.u. width up to 30 a.u. The panel at the lower right contains the events with the highest measured recoil-ion transversal momenta with 30 a.u. < $p_{r,rec}$ ≤ 40 a.u. The color scale used for all panels is shown in Fig. 1(b).

For nuclear momentum transfers above 12 a.u. the recoil-ion spectrometer does not have full solid angle of detection. Therefore the number of detected events at the different regions of $p_{r,rec}$ is not suitable to calculate the single differential cross section d$\sigma$/d$p_{r,rec}$.

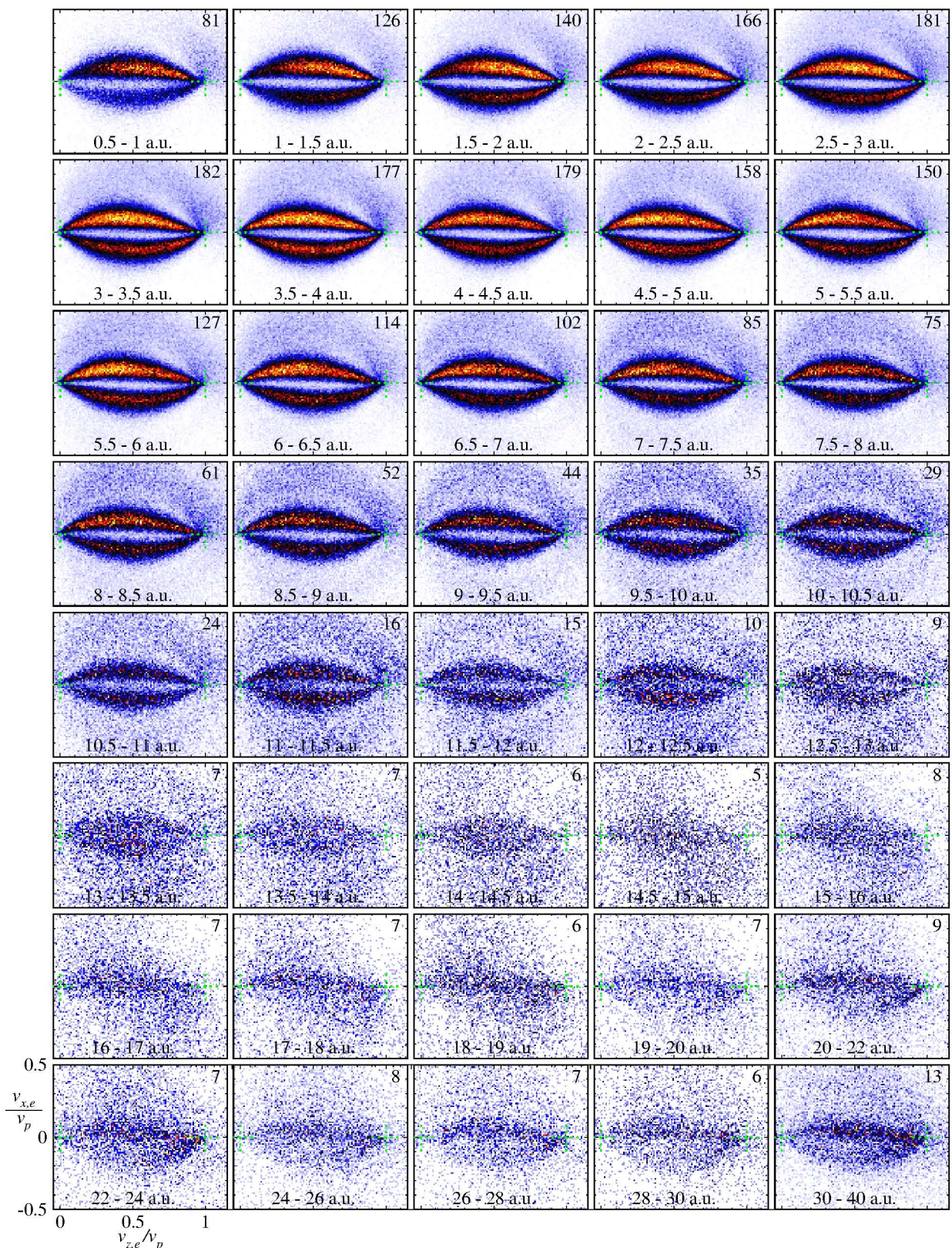

Fig. 2: Electron velocity distributions projected on the scattering plane (top view) for the transfer ionization in collisions of 15 keV/u $He^{2+}$ with He. See text for details.

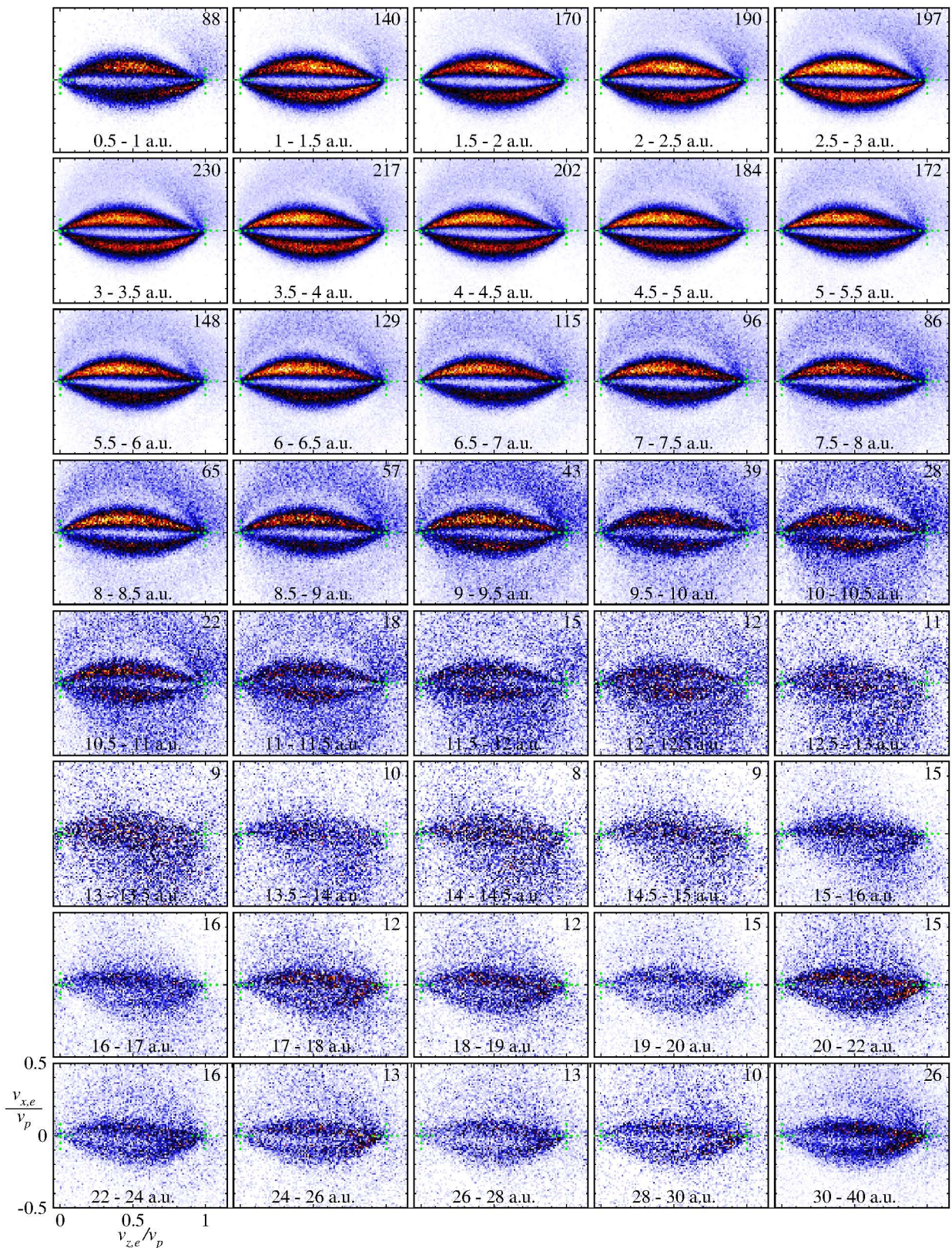

Fig. 3: Electron velocity distributions projected on the scattering plane (top view) for the transfer ionization in collisions of 13.3 keV/u $He^{2+}$ with He. See text for details.

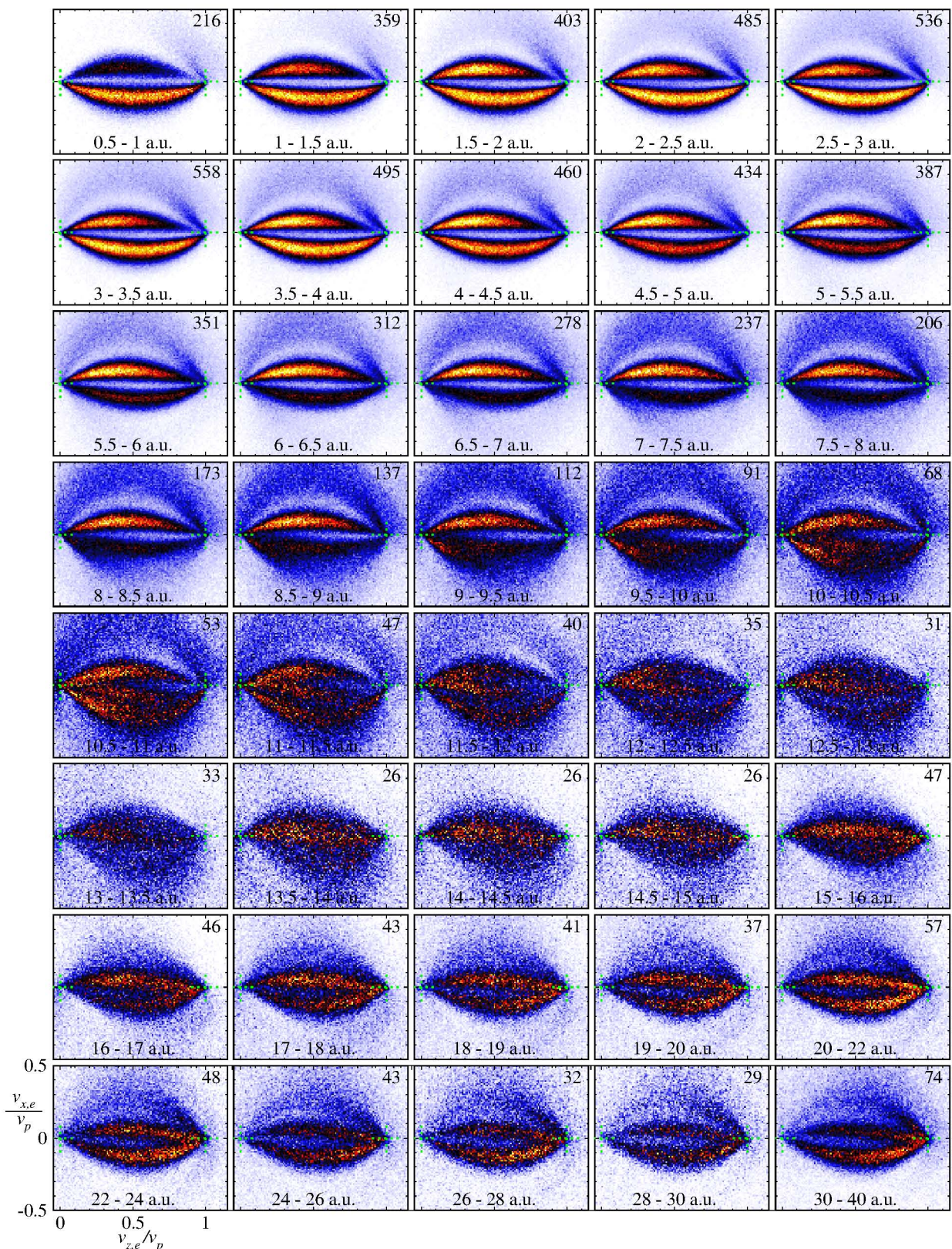

Fig. 4: Electron velocity distributions projected on the scattering plane (top view) for the transfer ionization in collisions of 11.5 keV/u $He^{2+}$ with He. See text for details.

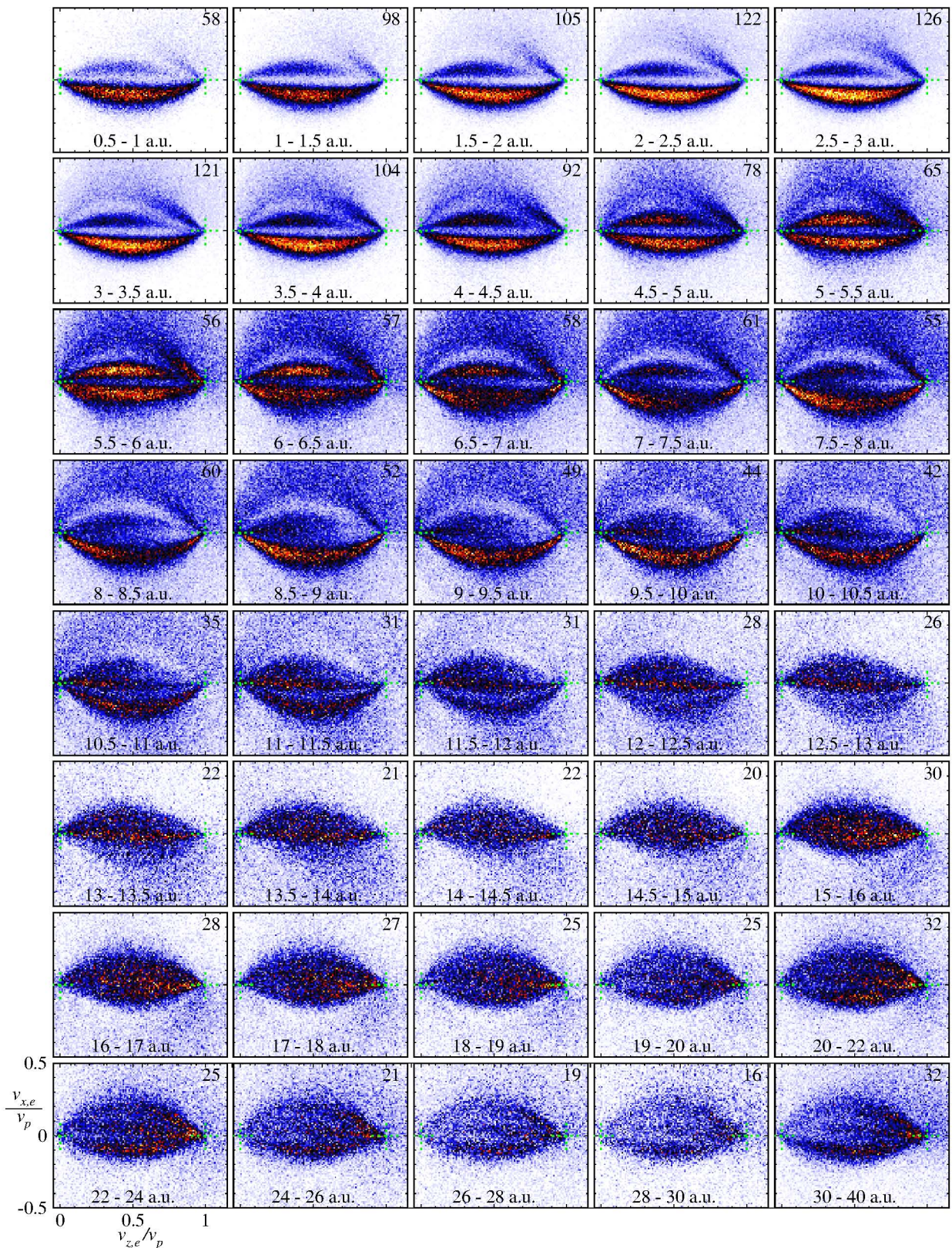

Fig. 5: Electron velocity distributions projected on the scattering plane (top view) for the transfer ionization in collisions of 10 keV/u He$^{2+}$ with He. See text for details.

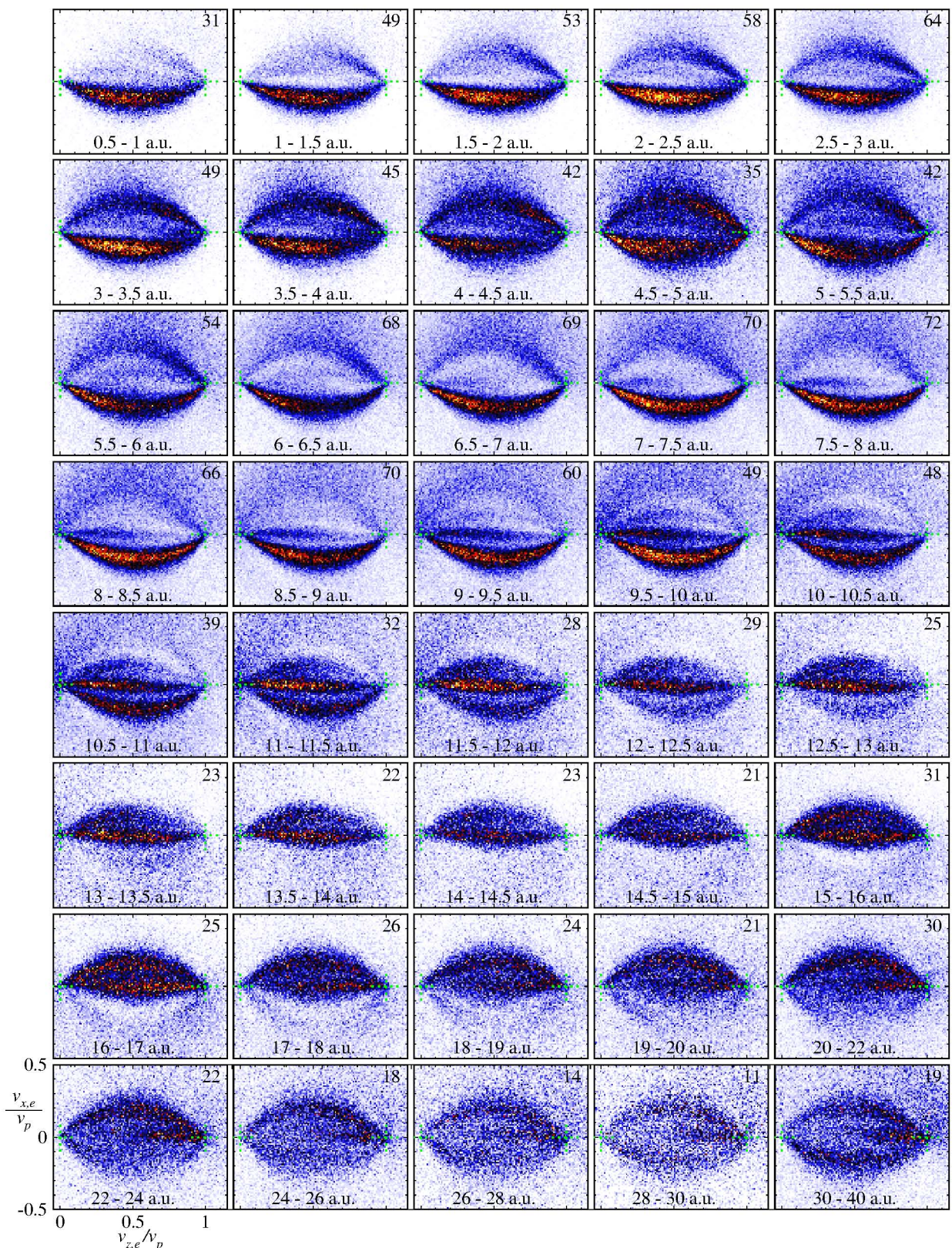

Fig. 6: Electron velocity distributions projected on the scattering plane (top view) for the transfer ionization in collisions of 8.5 keV/u He$^{2+}$ with He. See text for details.

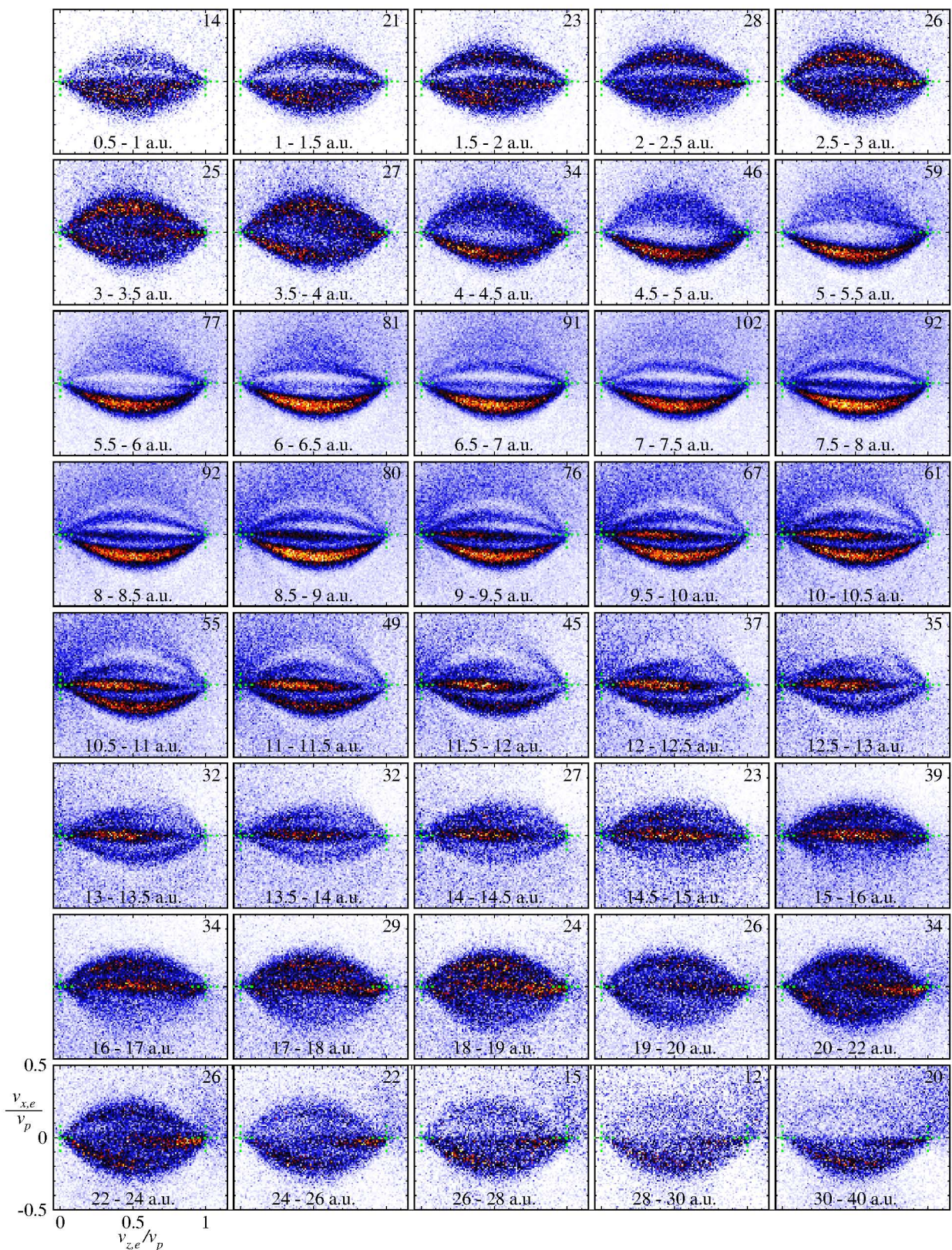

Fig. 7: Electron velocity distributions projected on the scattering plane (top view) for the transfer ionization in collisions of 7 keV/u $He^{2+}$ with He. See text for details.

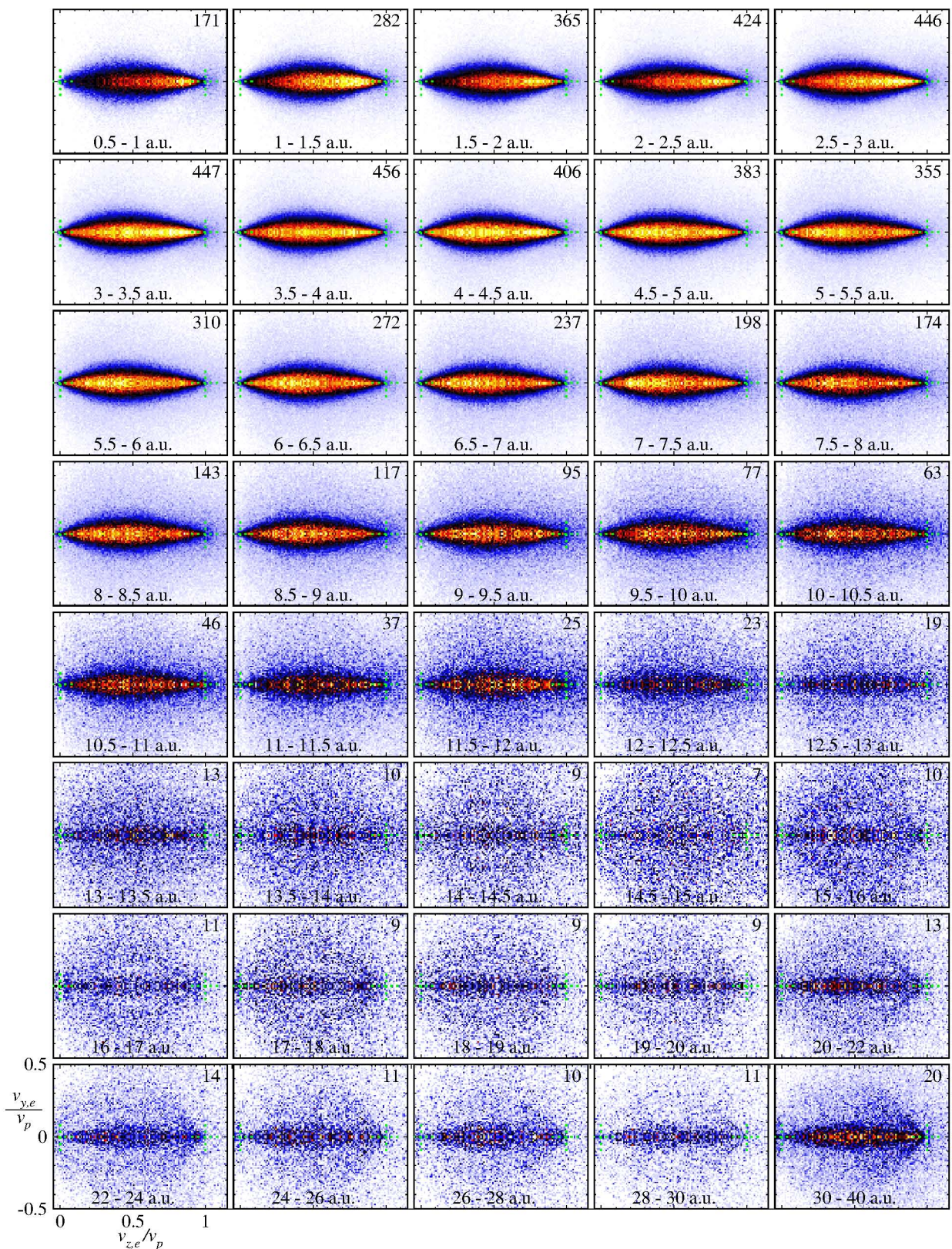

Fig. 8: Electron velocity distributions projected on the perpendicular plane (side view) for the transfer ionization in collisions of 15 keV/u He$^{2+}$ with He. See text for details.

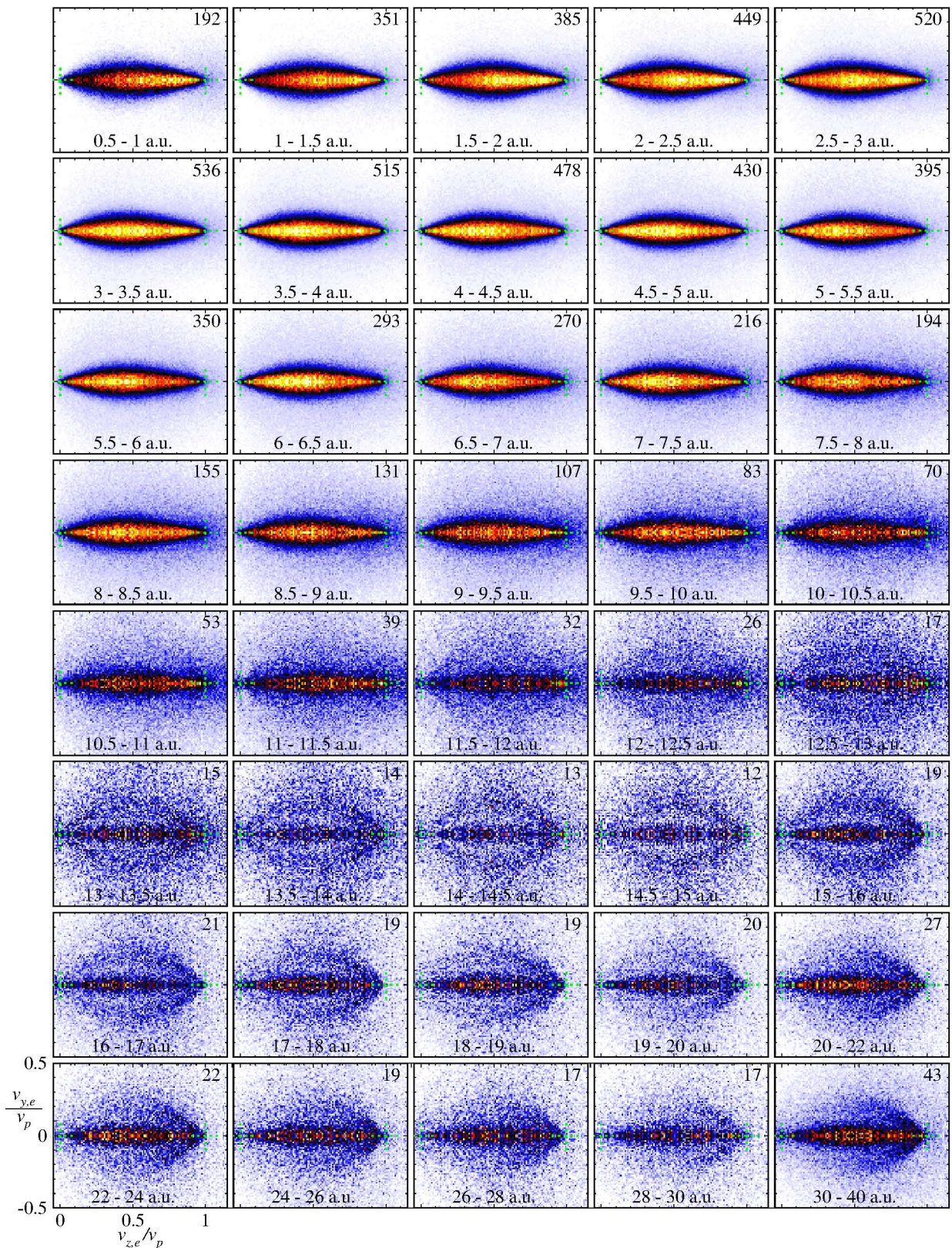

Fig. 9: Electron velocity distributions projected on the perpendicular plane (side view) for the transfer ionization in collisions of 13.3 keV/u He$^{2+}$ with He. See text for details.

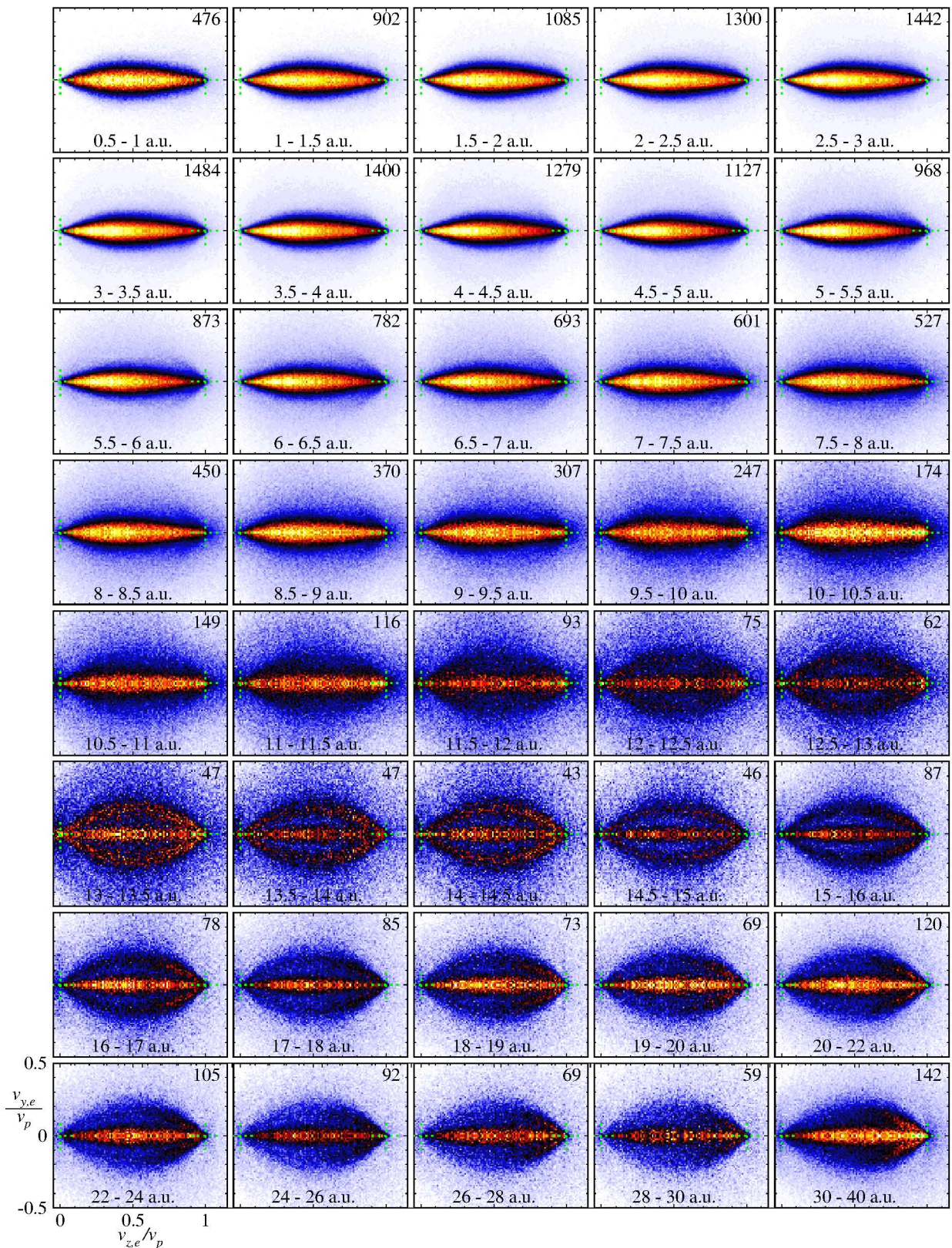

Fig. 10: Electron velocity distributions projected on the perpendicular plane (side view) for the transfer ionization in collisions of 11.5 keV/u He$^{2+}$ with He. See text for details.

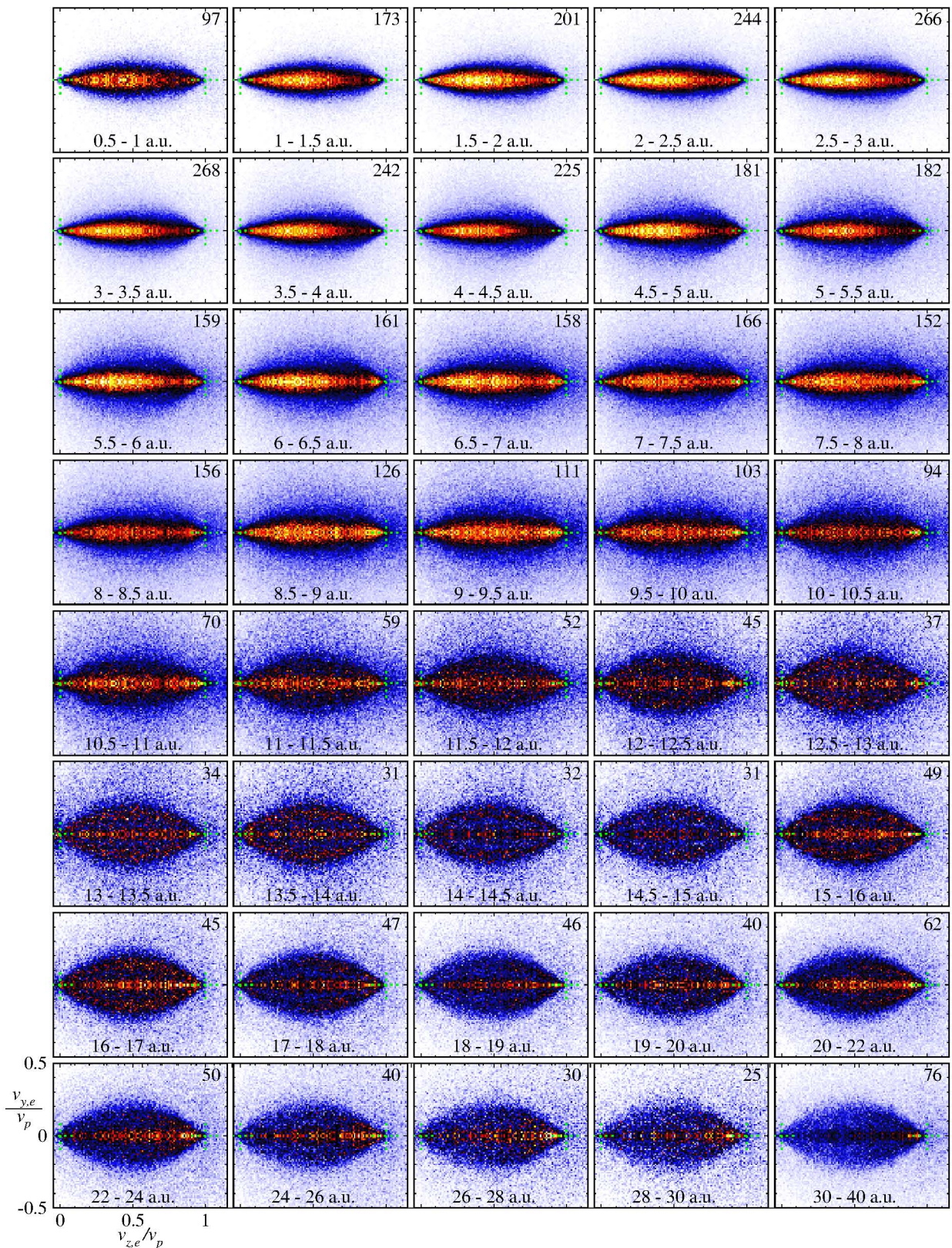

Fig. 11: Electron velocity distributions projected on the perpendicular plane (side view) for the transfer ionization in collisions of 10 keV/u He$^{2+}$ with He. See text for details.

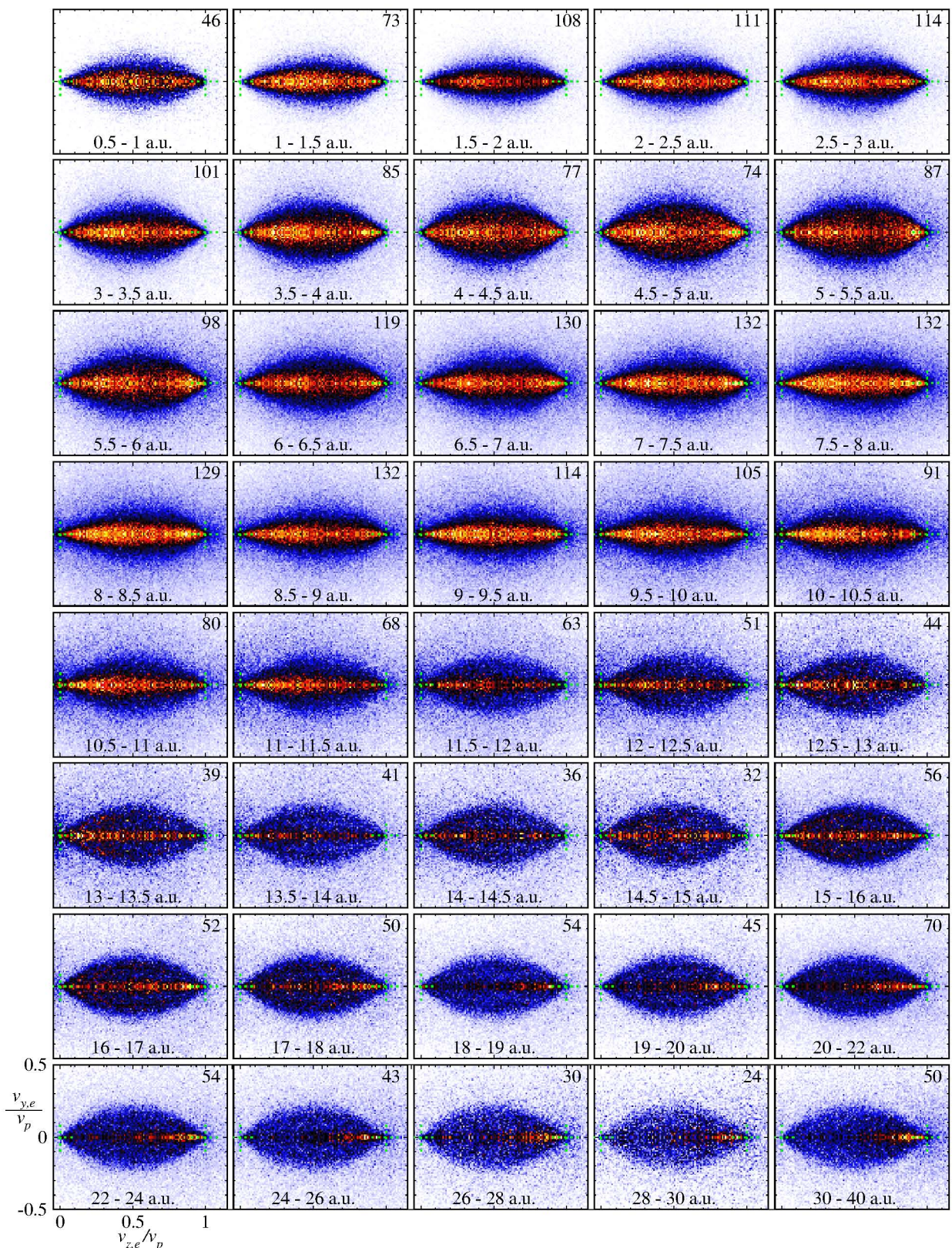

Fig. 12: Electron velocity distributions projected on the perpendicular plane (side view) for the transfer ionization in collisions of 8.5 keV/u He$^{2+}$ with He. See text for details.

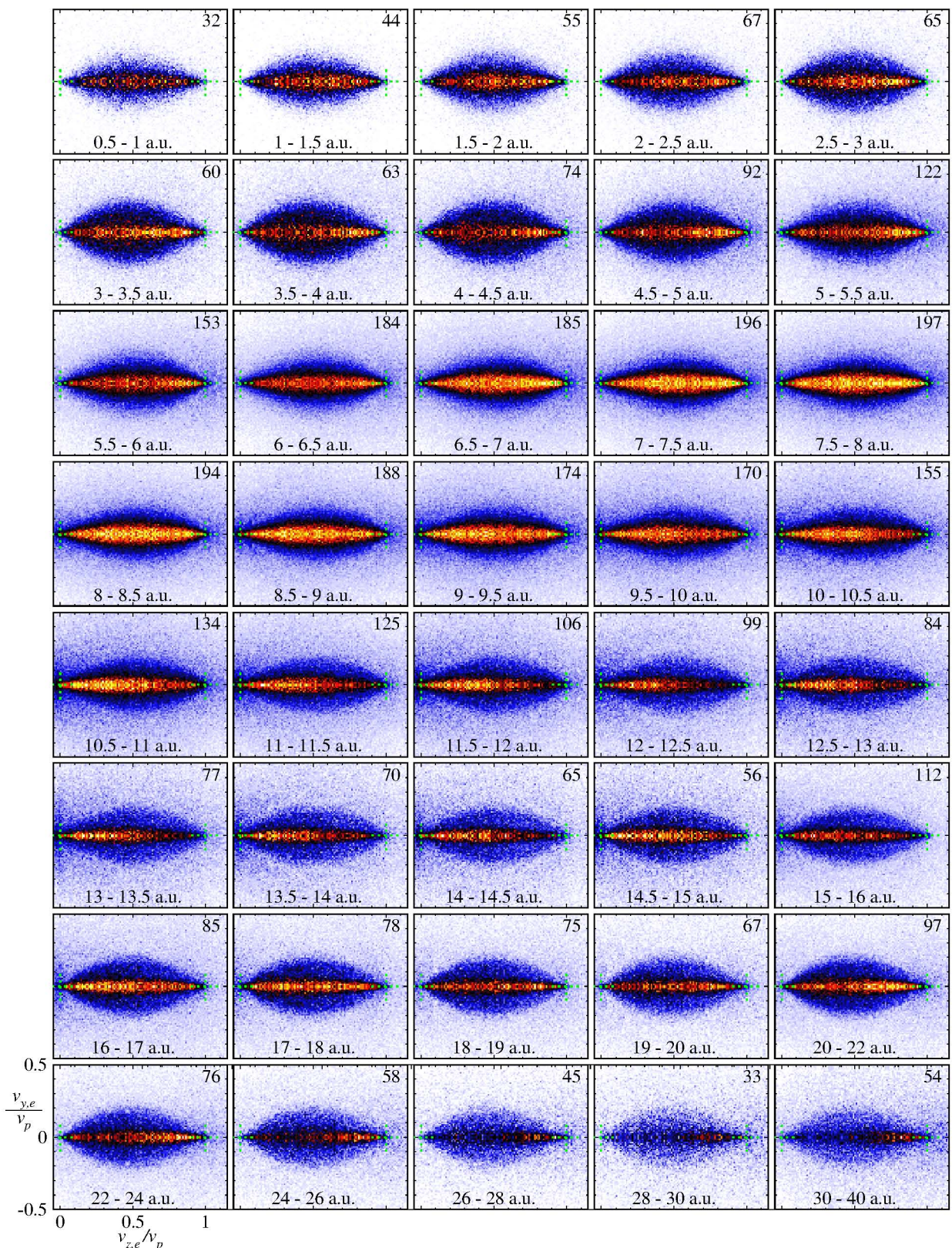

Fig. 13: Electron velocity distributions projected on the perpendicular plane (side view) for the transfer ionization in collisions of 7 keV/u $He^{2+}$ with He. See text for details.

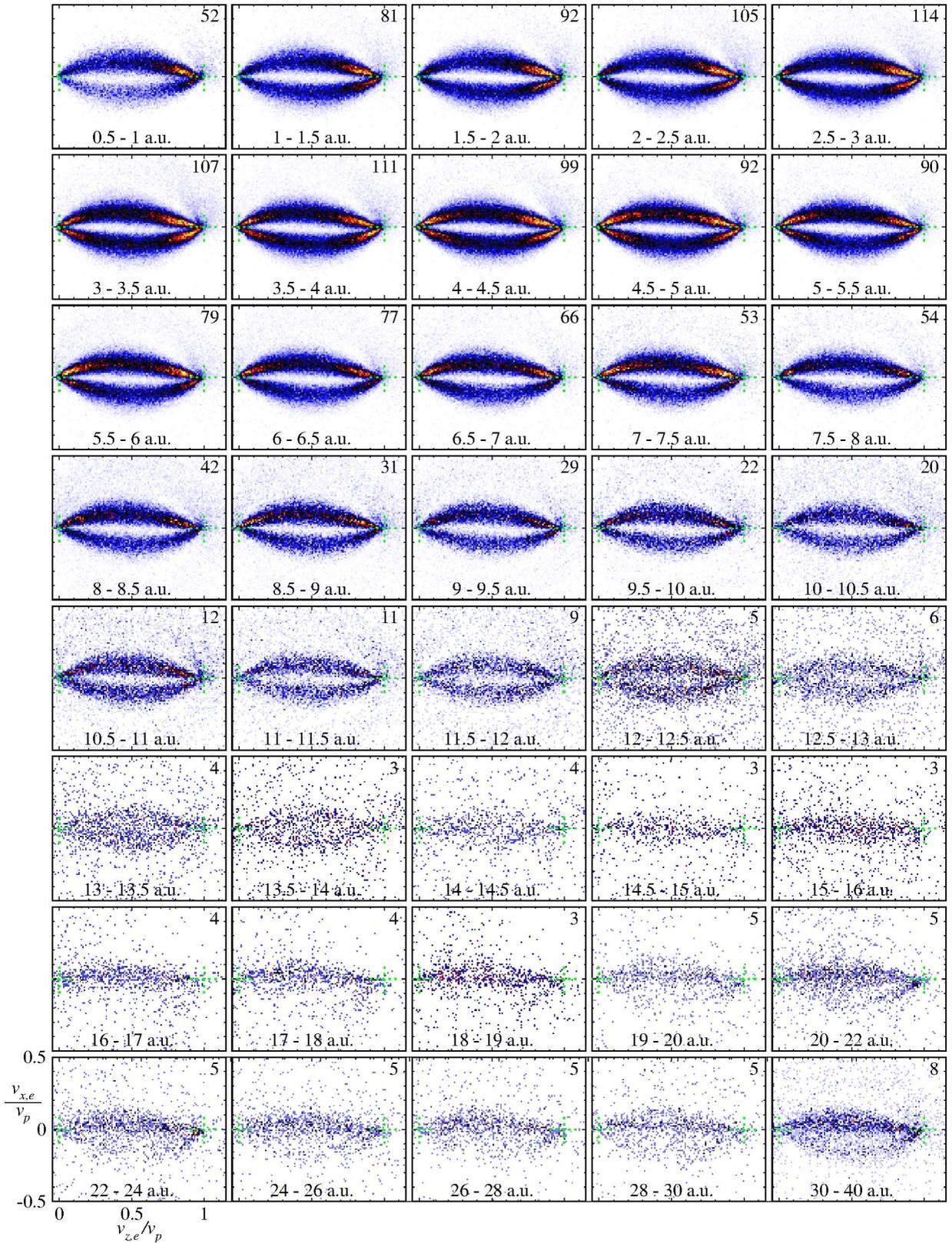

Fig. 14: Scattering plane electron velocity distributions for the transfer ionization in collisions of 15 keV/u $He^{2+}$ with He. See text for details.

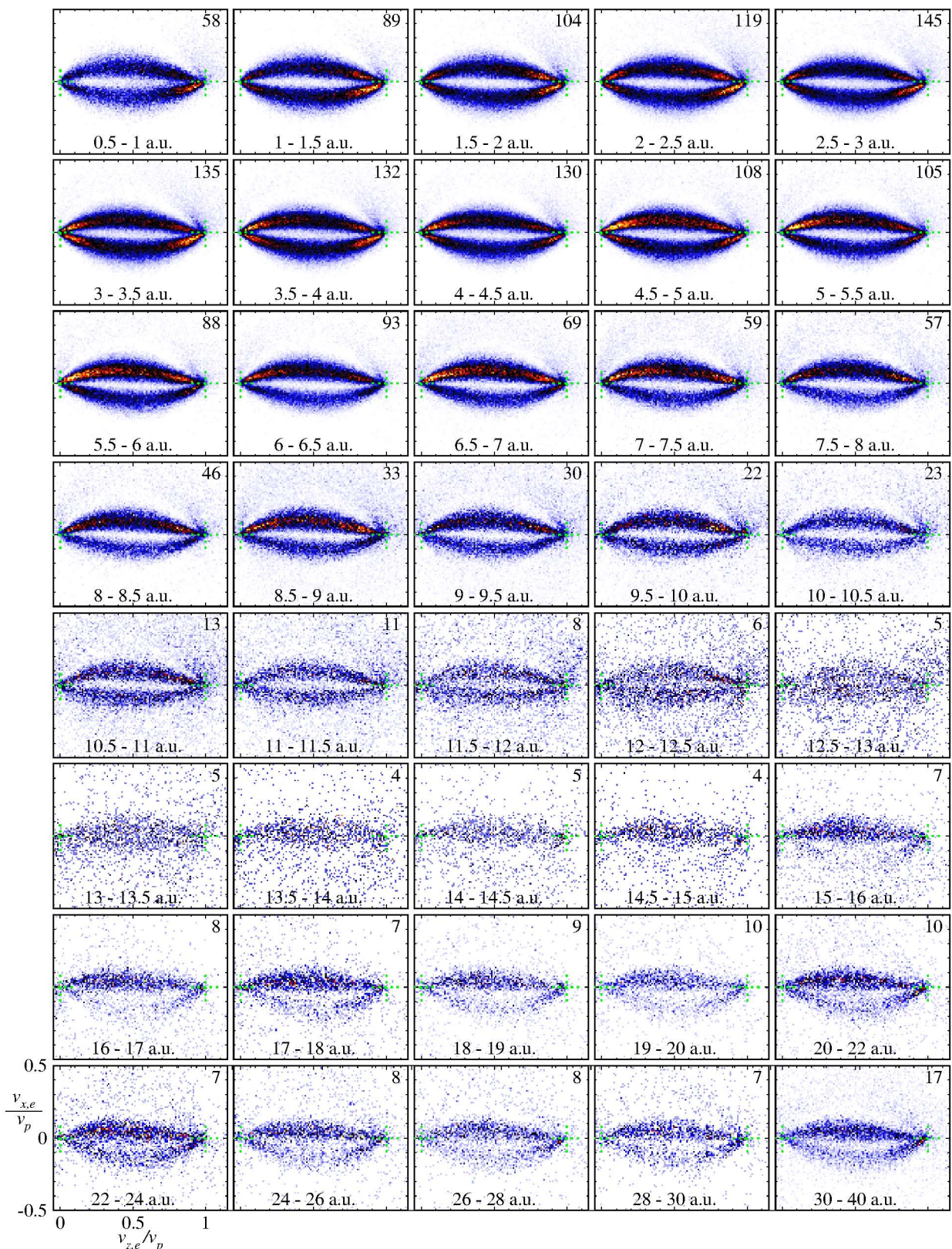

Fig. 15: Scattering plane electron velocity distributions for the transfer ionization in collisions of 13.3 keV/u He$^{2+}$ with He. See text for details.

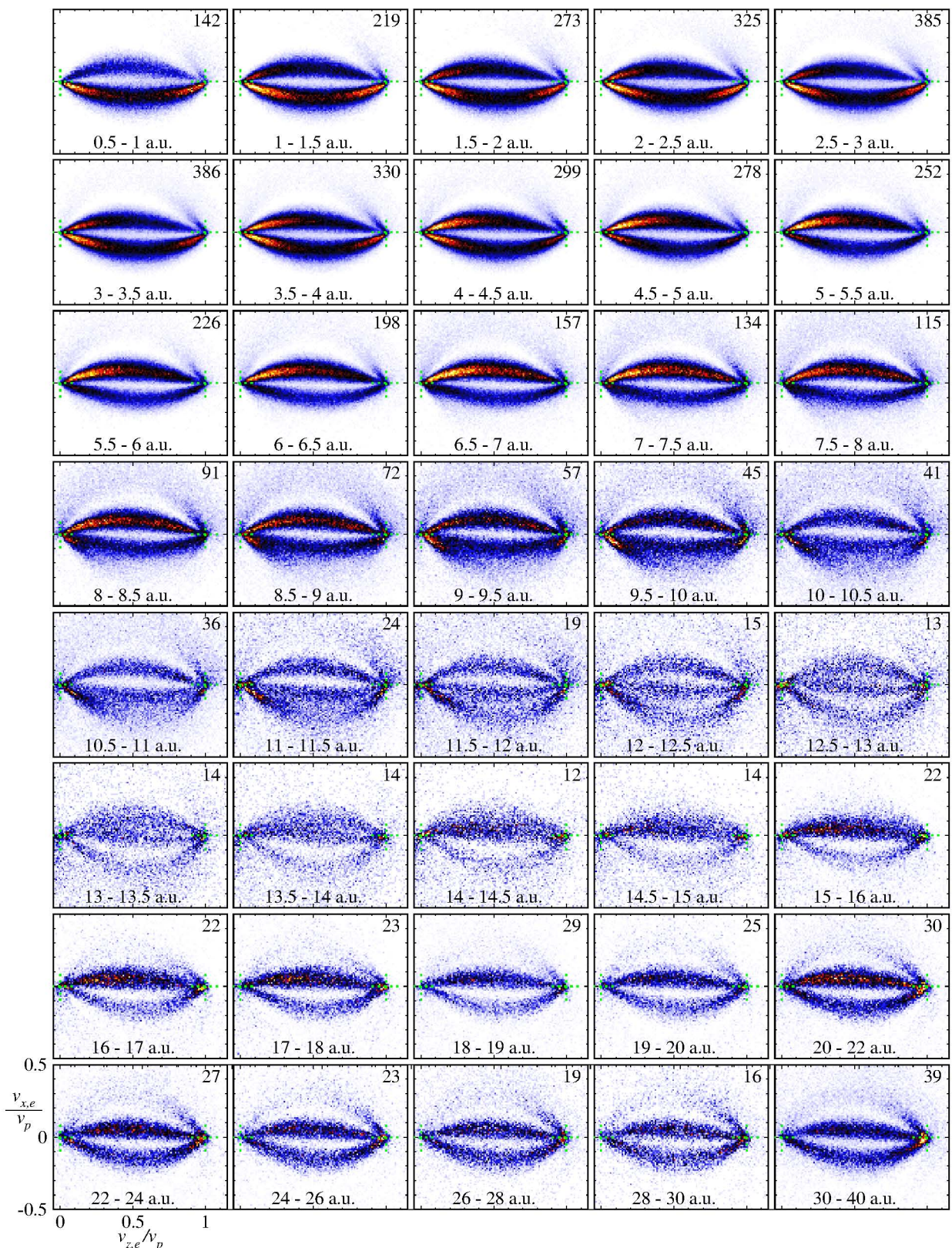

Fig. 16: Scattering plane electron velocity distributions for the transfer ionization in collisions of 11.5 keV/u $He^{2+}$ with He. See text for details.

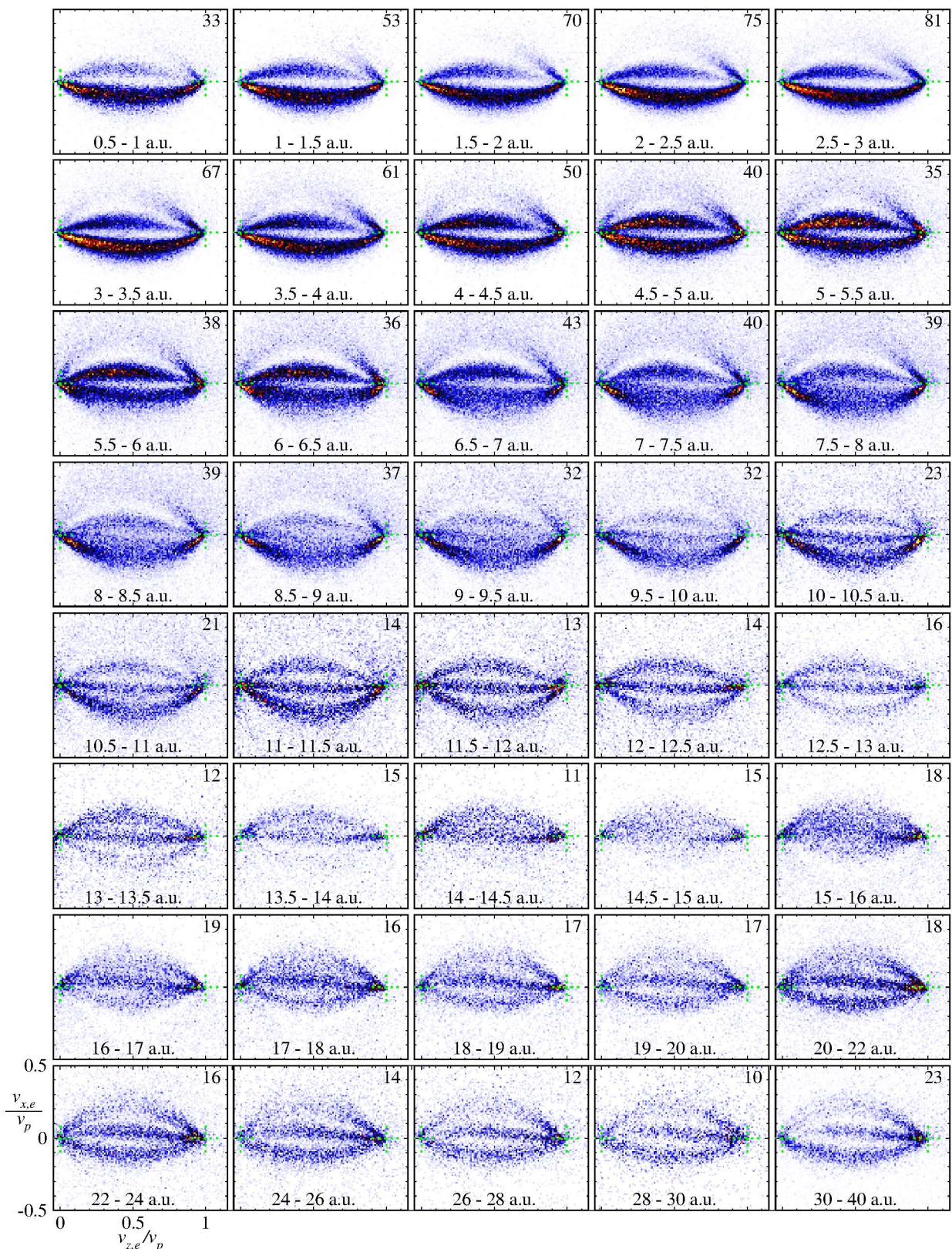

Fig. 17: Scattering plane electron velocity distributions for the transfer ionization in collisions of 10 keV/u He$^{2+}$ with He. See text for details.

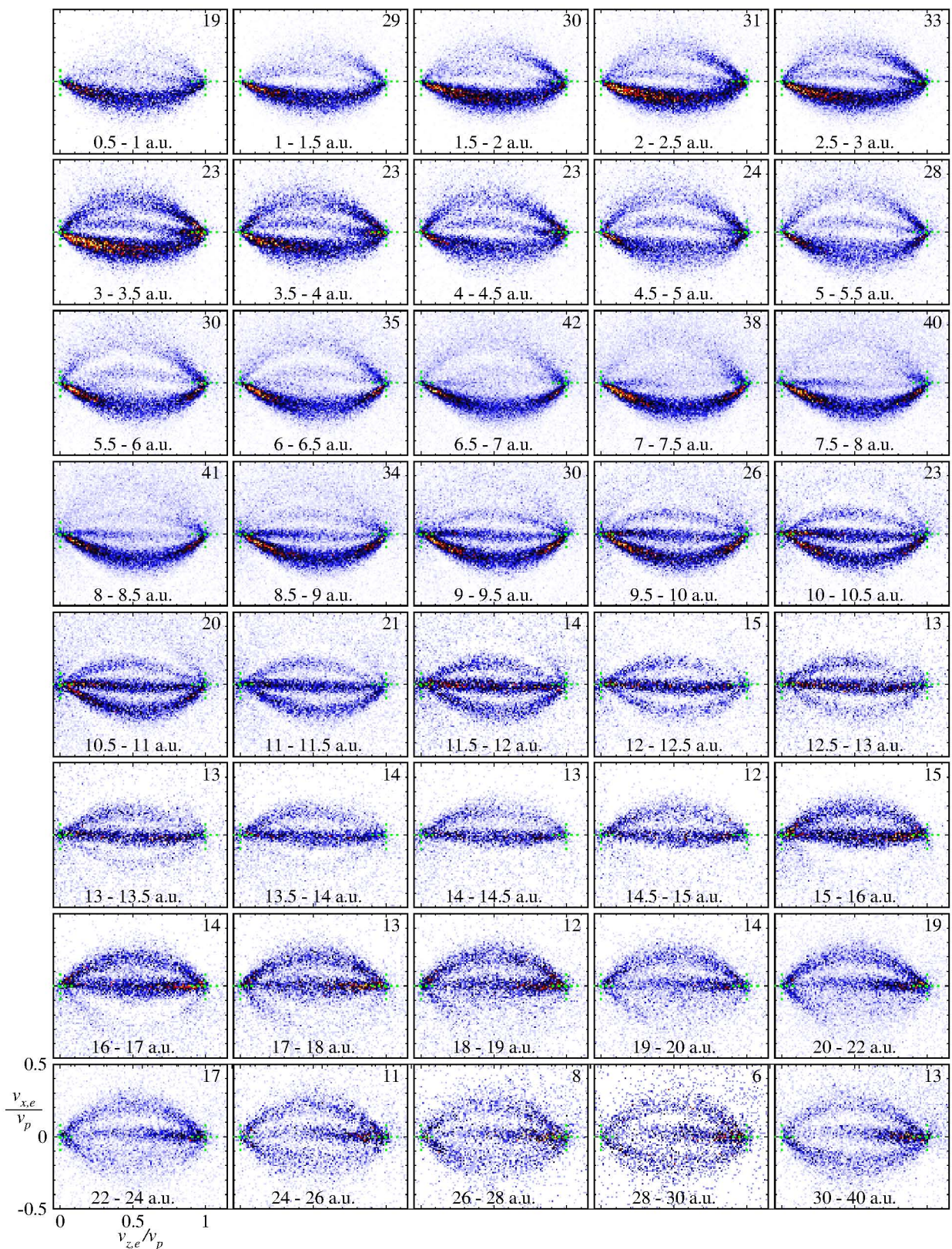

Fig. 18: Scattering plane electron velocity distributions for the transfer ionization in collisions of 8.5 keV/u $He^{2+}$ with He. See text for details.

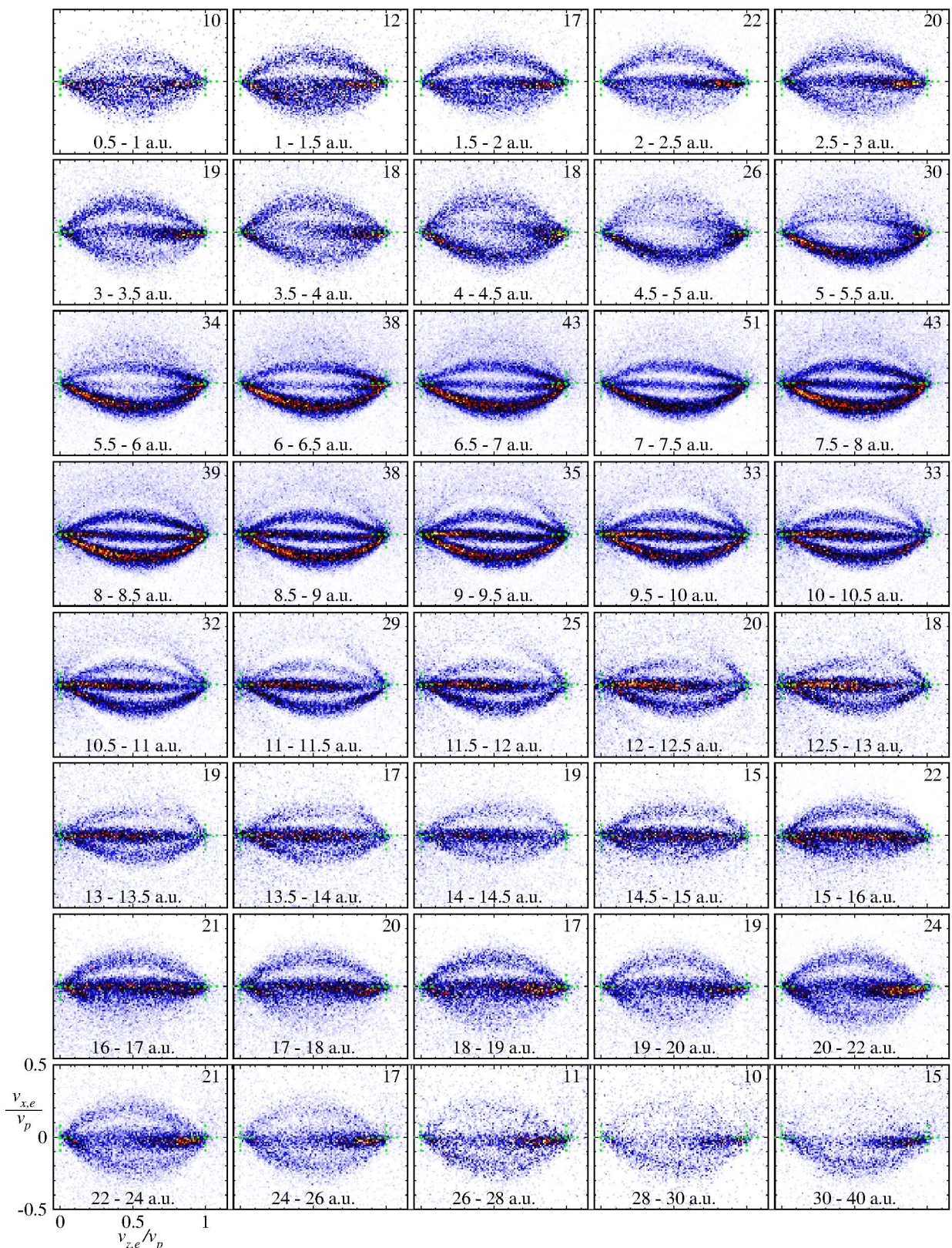

Fig. 19: Scattering plane electron velocity distributions for the transfer ionization in collisions of 7 keV/u He$^{2+}$ with He. See text for details.

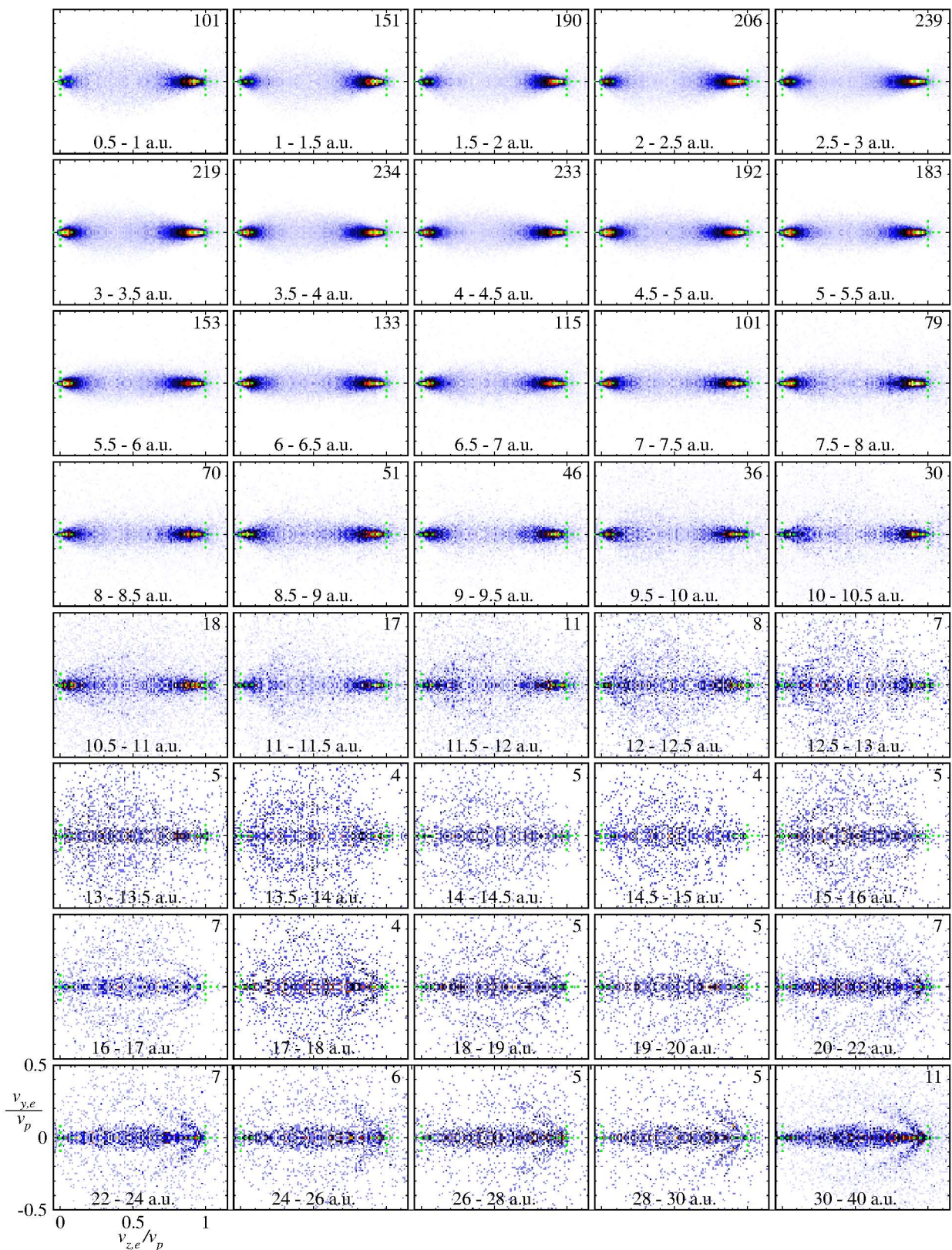

Fig. 20: Perpendicular plane electron velocity distributions for the transfer ionization in collisions of 15 keV/u He$^{2+}$ with He. See text for details.

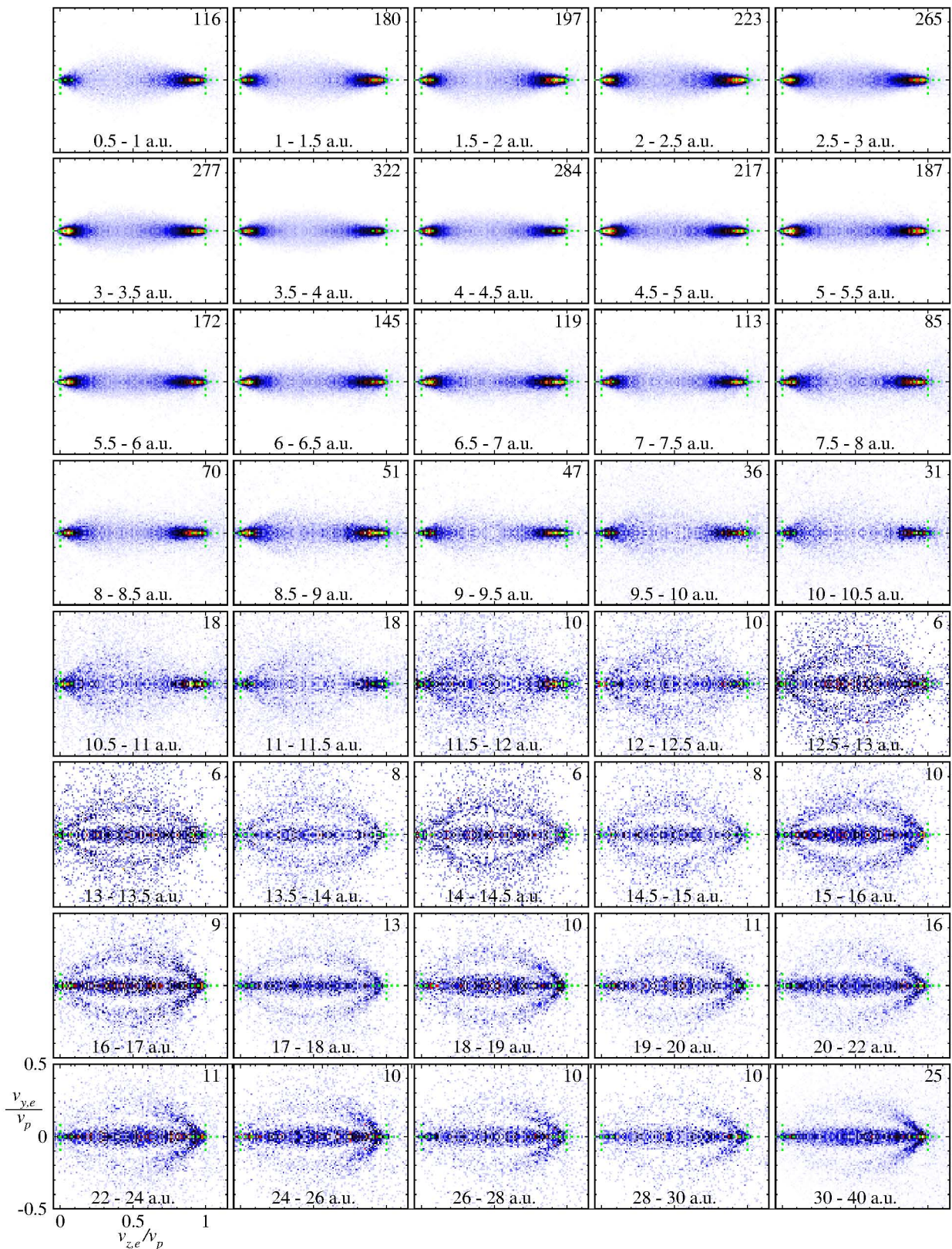

Fig. 21: Perpendicular plane electron velocity distributions for the transfer ionization in collisions of 13.3 keV/u $He^{2+}$ with He. See text for details.

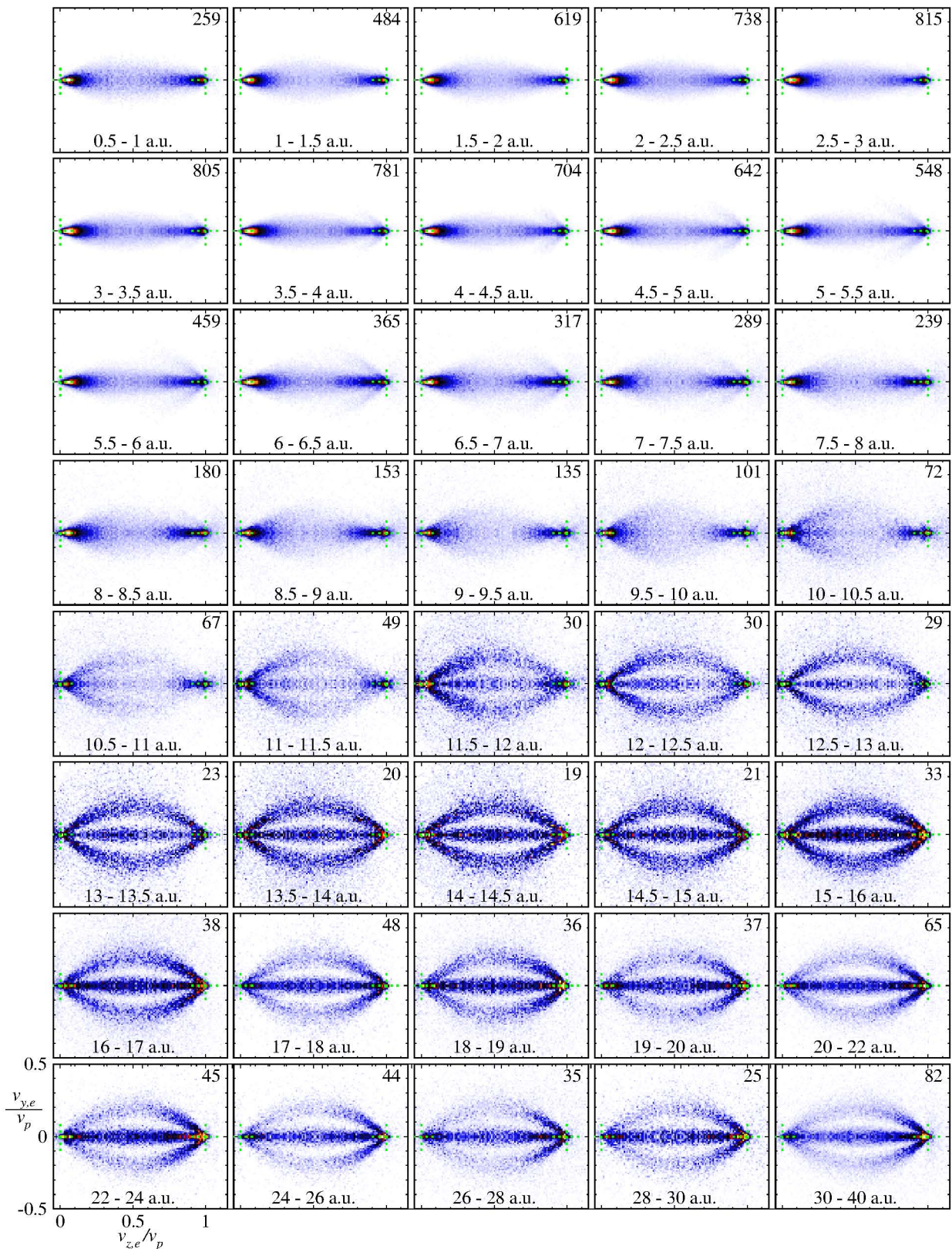

Fig. 22: Perpendicular plane electron velocity distributions for the transfer ionization in collisions of 11.5 keV/u He$^{2+}$ with He. See text for details.

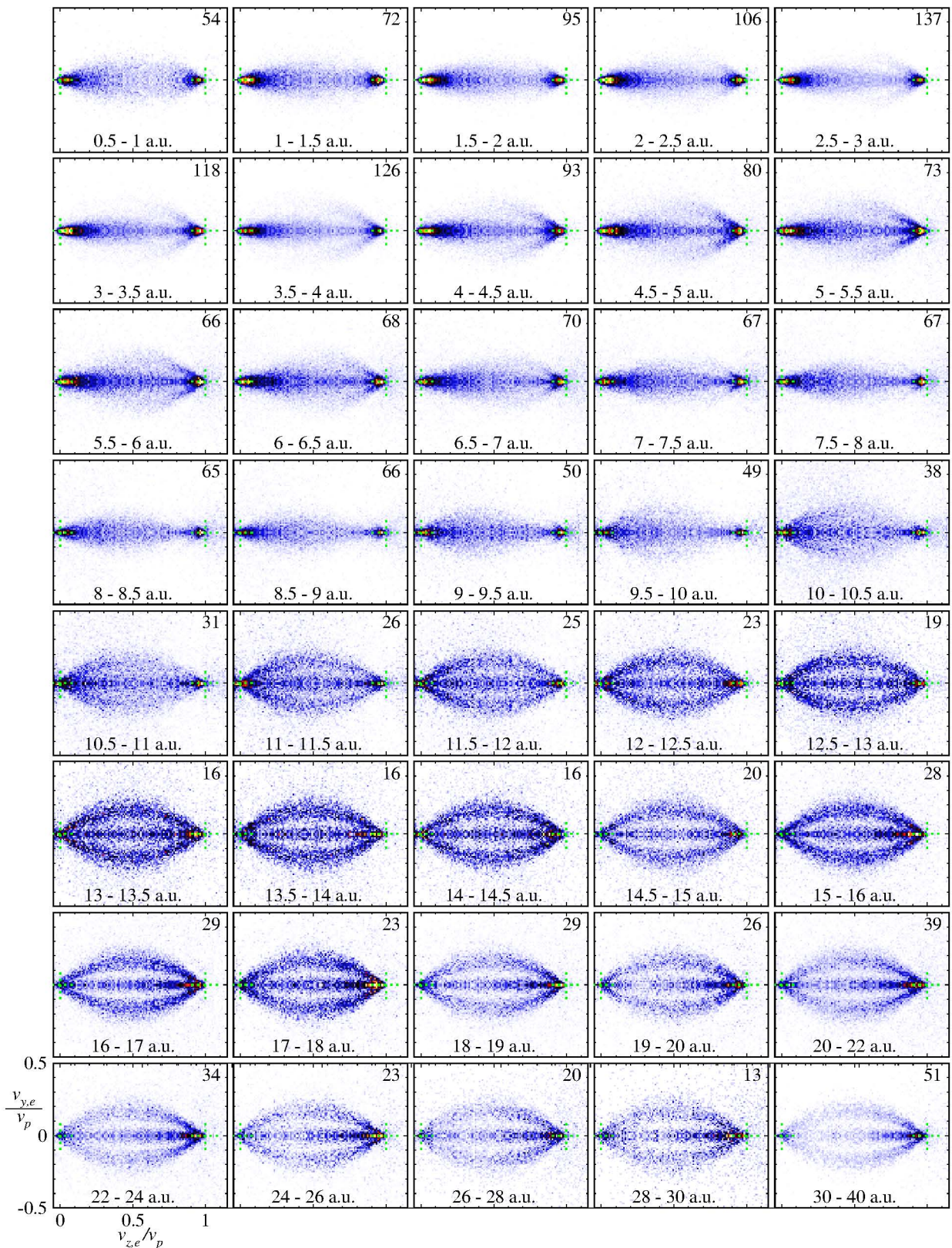

Fig. 23: Perpendicular plane electron velocity distributions for the transfer ionization in collisions of 10 keV/u $He^{2+}$ with He. See text for details.

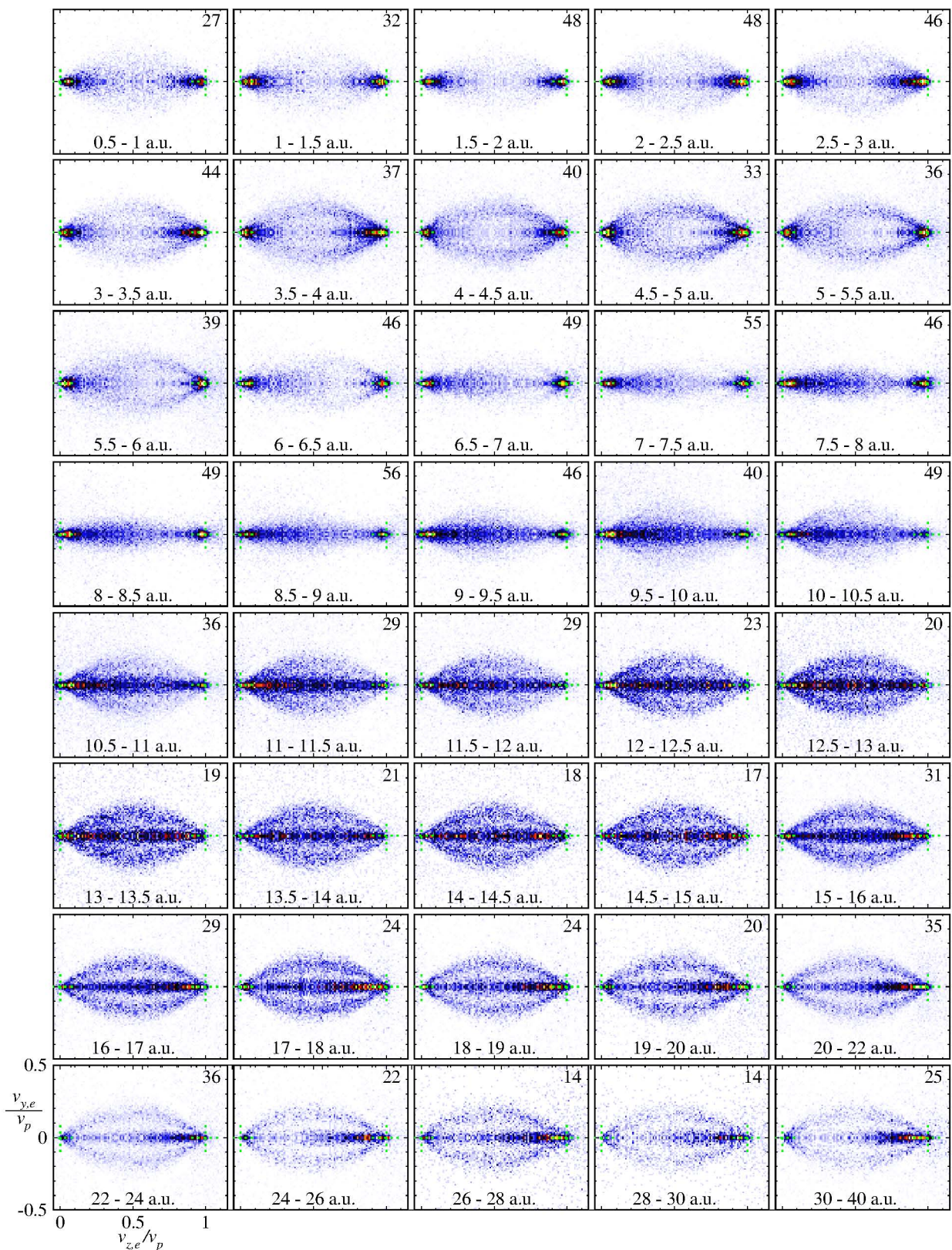

Fig. 24: Perpendicular plane electron velocity distributions for the transfer ionization in collisions of 8.5 keV/u $He^{2+}$ with He. See text for details.

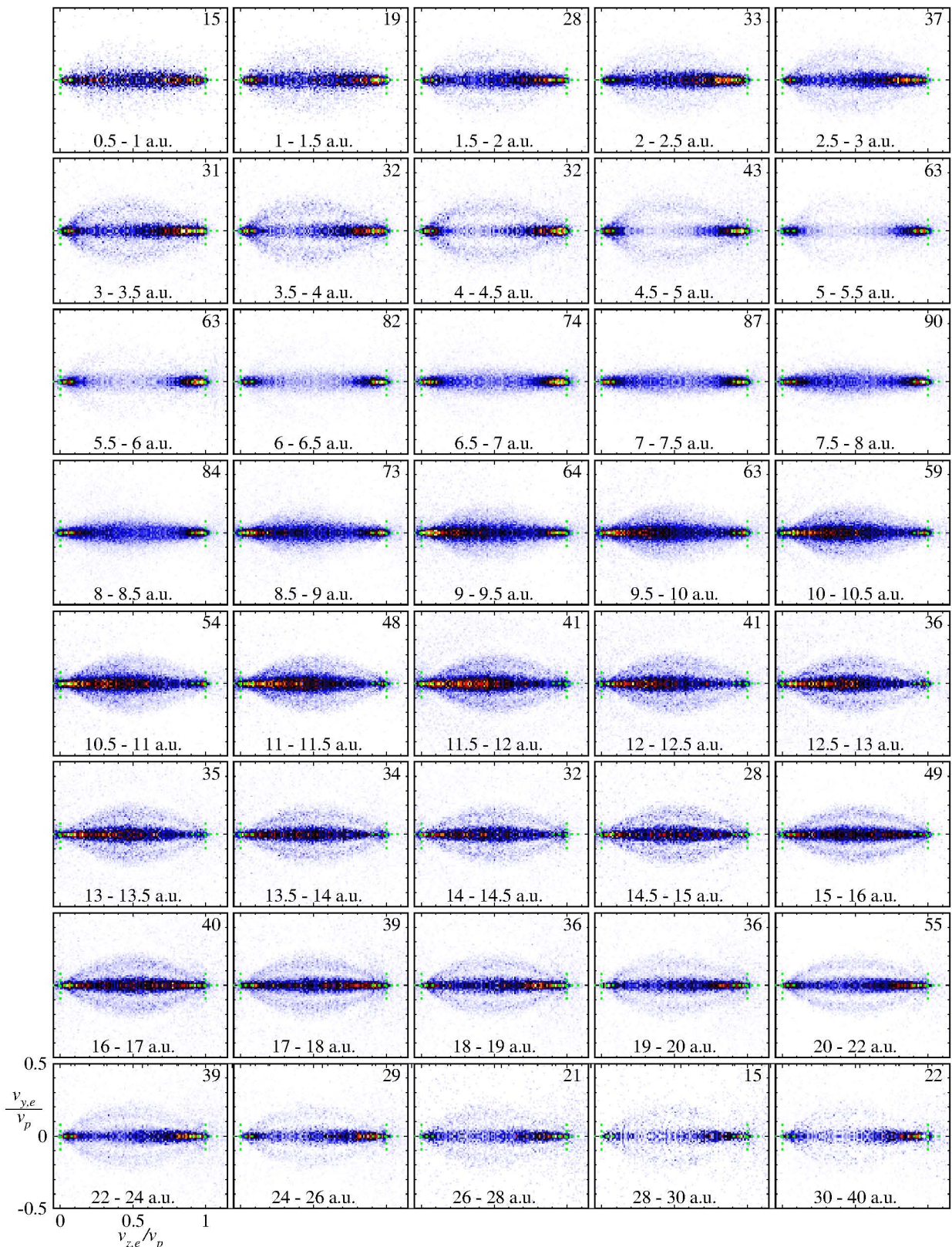

Fig. 25: Perpendicular plane electron velocity distributions for the transfer ionization in collisions of 7 keV/u $He^{2+}$ with He. See text for details.

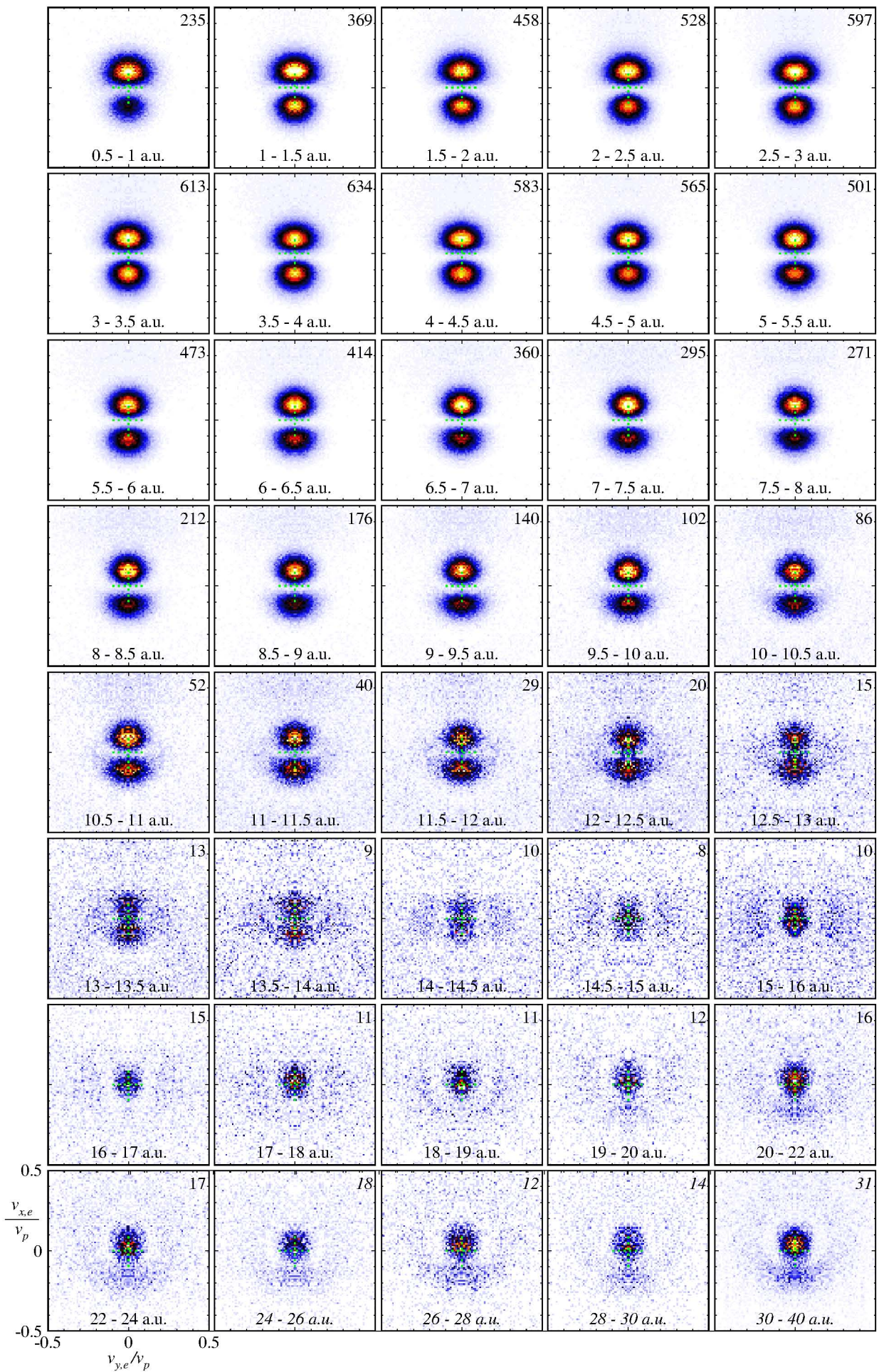

Fig. 26: Transverse plane scaled electron velocity distributions for the transfer ionization in collisions of 15 keV/u $He^{2+}$ with He. See text for details.

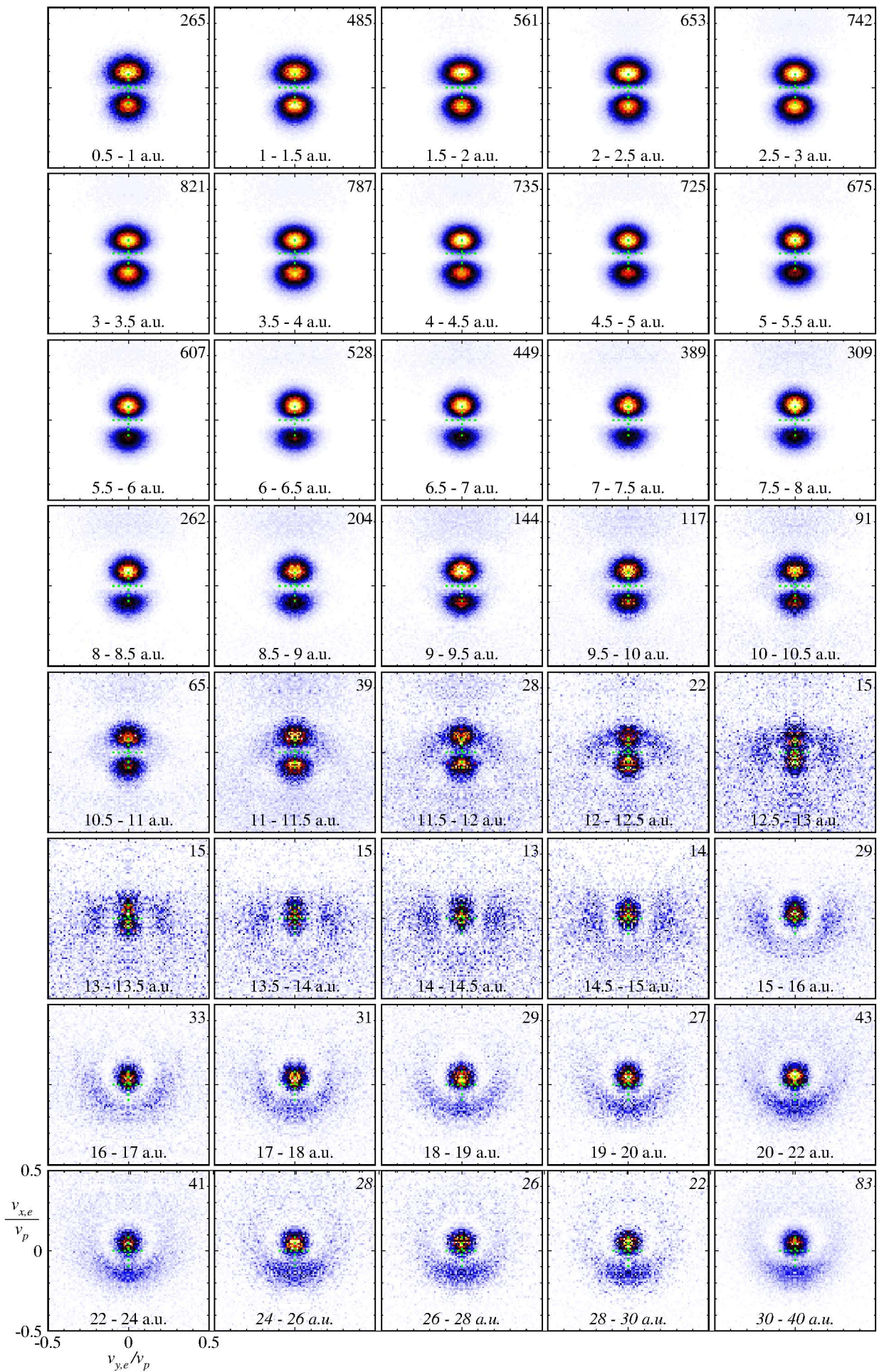

Fig. 27: Transverse plane scaled electron velocity distributions for the transfer ionization in collisions of 13.3 keV/u $He^{2+}$ with He. See text for details.

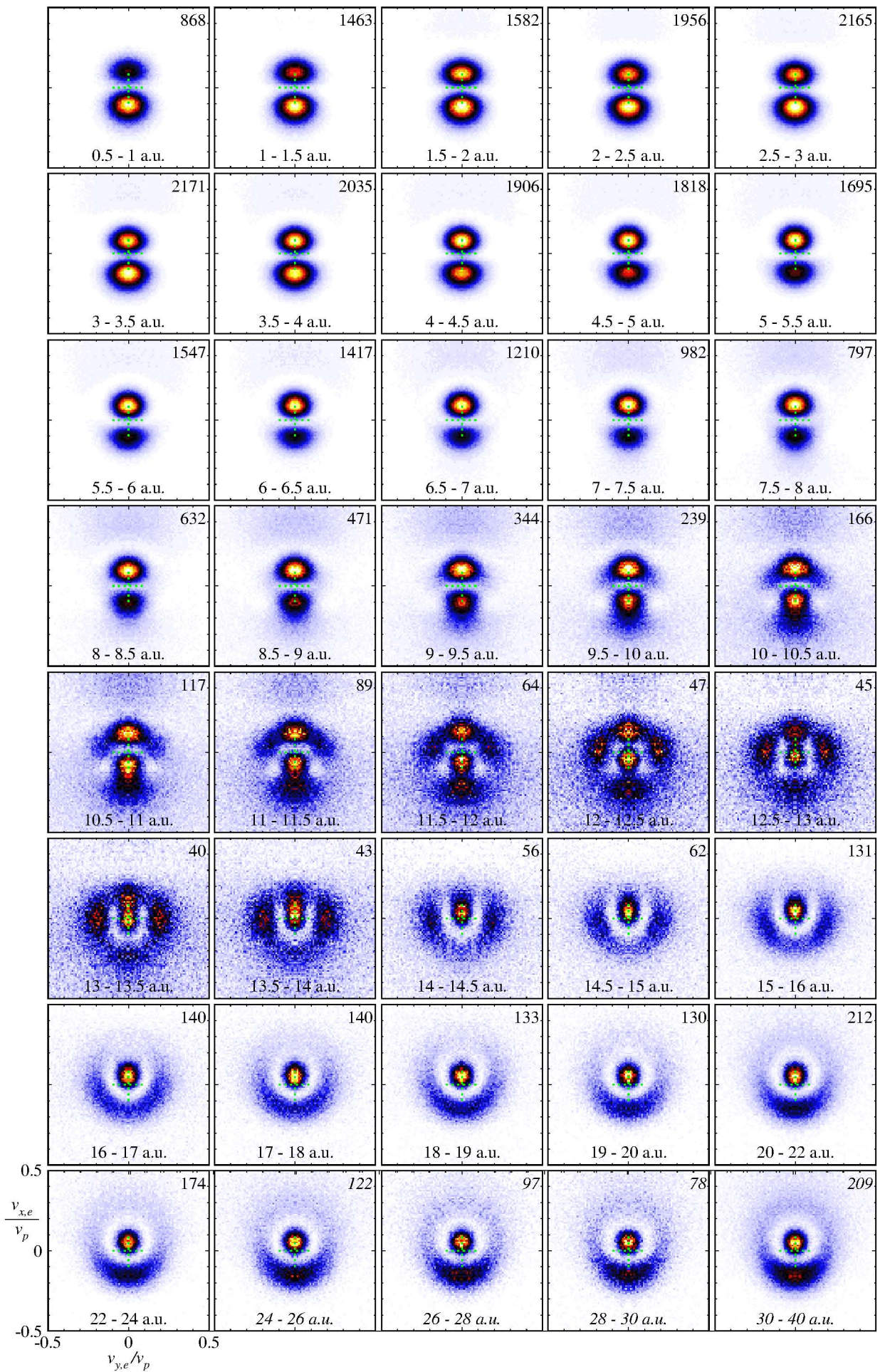

Fig. 28: Transverse plane scaled electron velocity distributions for the transfer ionization in collisions of 11.5 keV/u $He^{2+}$ with He. See text for details.

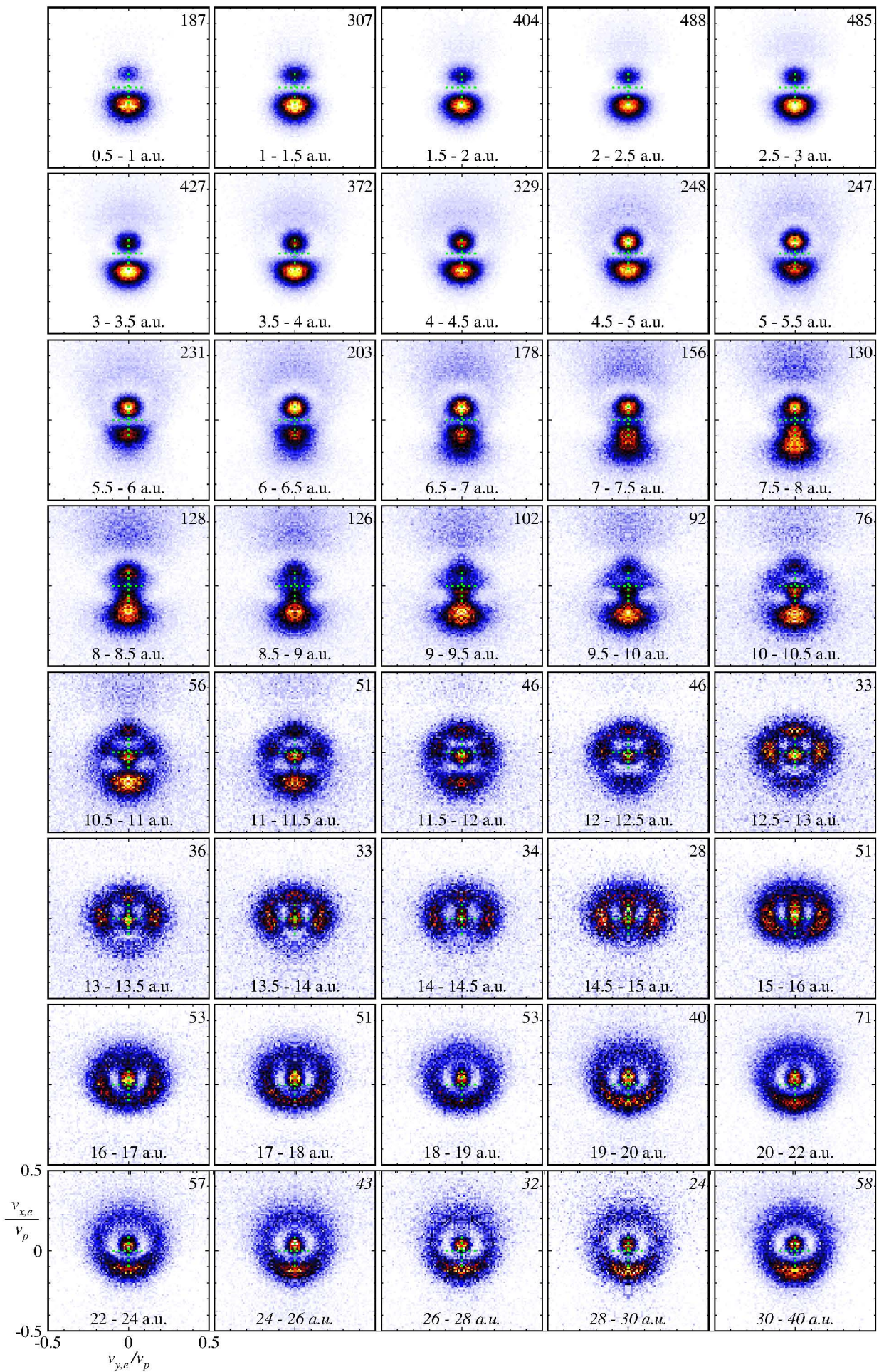

Fig. 29: Transverse plane scaled electron velocity distributions for the transfer ionization in collisions of 10 keV/u He$^{2+}$ with He. See text for details.

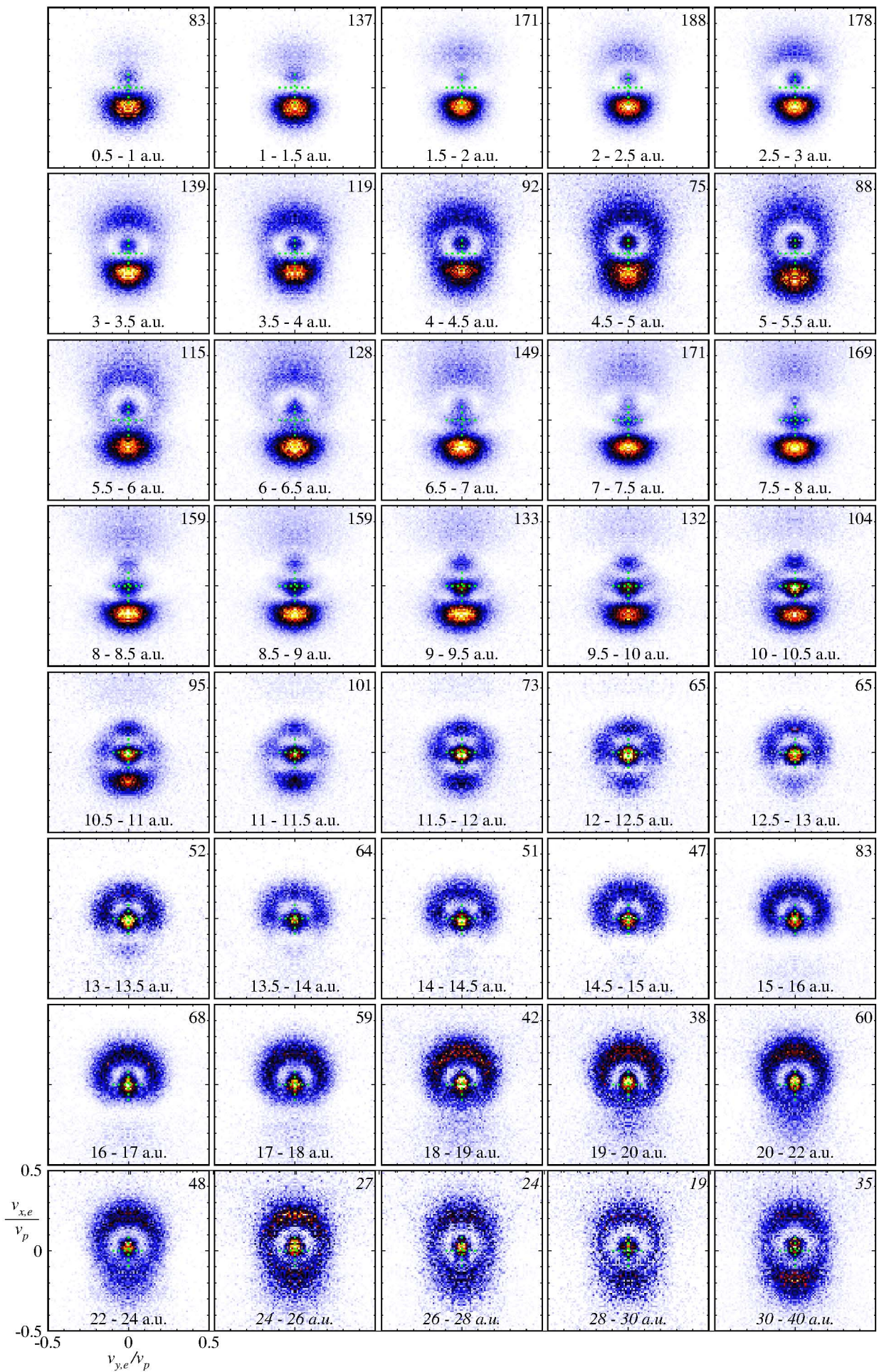

Fig. 30: Transverse plane scaled electron velocity distributions for the transfer ionization in collisions of 8.5 keV/u He$^{2+}$ with He. See text for details.

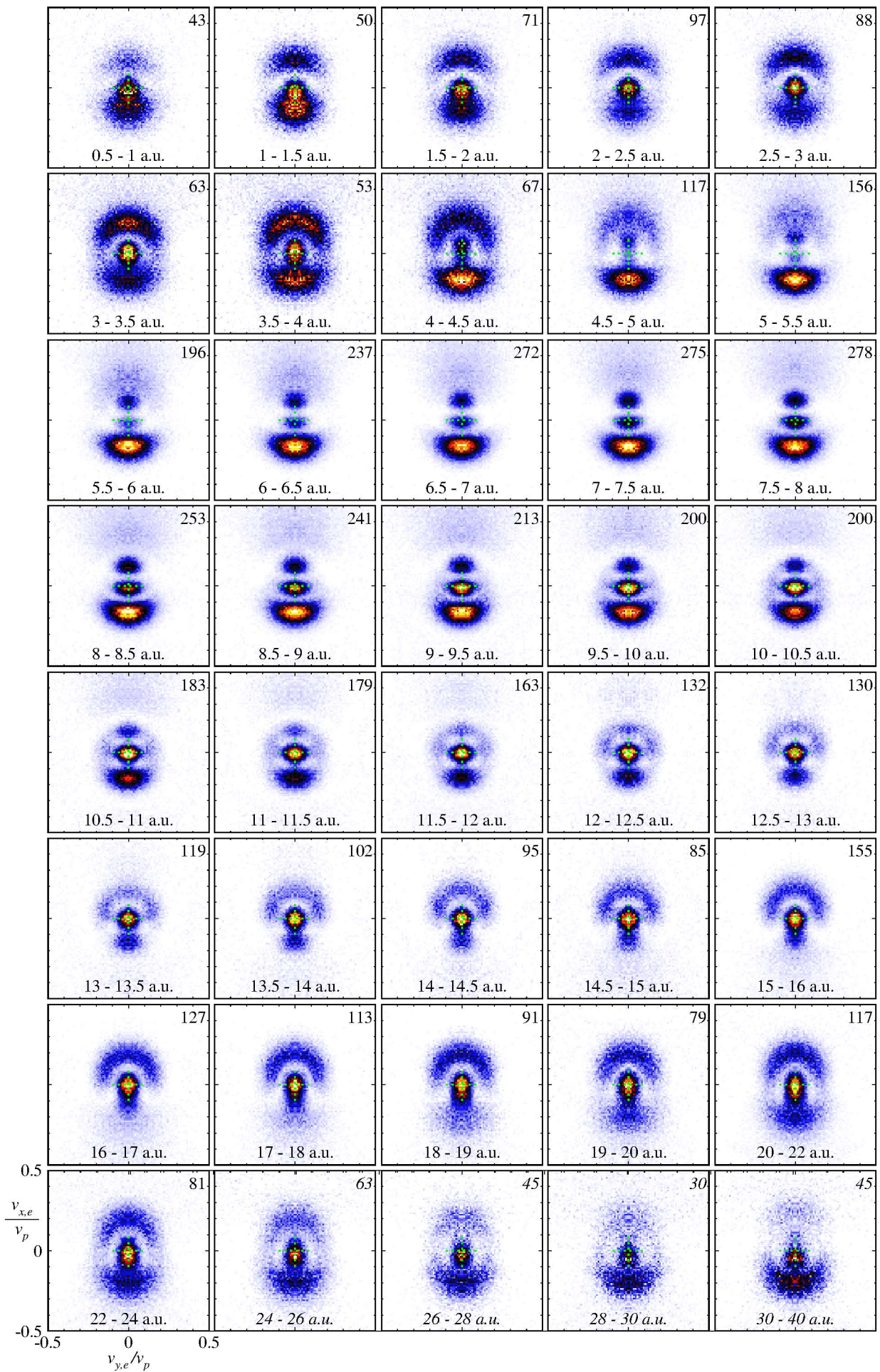

Fig. 31: Transverse plane scaled electron velocity distributions for the transfer ionization in collisions of 7 keV/u $He^{2+}$ with He. See text for details.